\documentclass[useAMS,usenatbib]{mn2e}

\DeclareSymbolFont{cmletters}{OML}{cmm}{m}{it}
\DeclareMathSymbol{v}{\mathalpha}{cmletters}{"76}

\usepackage{tabularx,ragged2e,booktabs,caption}
\usepackage{amsmath}
\usepackage{amssymb}
\usepackage{epsfig}
\usepackage{graphicx}
\usepackage{ifthen}
\usepackage{latexsym}
\usepackage{rotating}
\usepackage{subfigure}
\usepackage{times,epsf}
\usepackage{txfonts}
\usepackage{varioref}
\usepackage{verbatim}
\usepackage{array}
\usepackage{url}
\usepackage{color}
\usepackage[dvipsnames]{xcolor}
\usepackage[T1]{fontenc}

\newcolumntype{L}[1]{>{\raggedright\arraybackslash}p{#1}}
\newcolumntype{C}[1]{>{\centering\arraybackslash}p{#1}}
\newcolumntype{R}[1]{>{\raggedleft\arraybackslash}p{#1}}

\newcommand{\be}{\begin{equation}}
\newcommand{\ee}{\end{equation}}
\newcommand{\bea}{\begin{eqnarray}}
\newcommand{\eea}{\end{eqnarray}}

\newcommand\apj{Astrophysical Journal}
\newcommand\apjl{Astrophysical Journal Letters}

\newcommand\aap{Astronomy \& Astrophysics}

\newcommand\mnras{Monthly Notices of the Royal Astronomical Society}

\newcommand{\avg}[1]{\langle #1\rangle}

\newcommand{\koral}{\texttt{KORAL}\,}
\newcommand{\Medd}{\dot M_{\rm Edd}}
\newcommand{\ledd}{L_{\rm Edd}}
\newcommand{\Ledd}{L_{\rm Edd}}
\newcommand{\medd}{\dot M_{\rm Edd}}
\newcommand{\Msun}{M_\odot}
\newcommand{\msun}{M_\odot}
\newcommand{\Rsun}{R_\odot}
\newcommand{\rg}{r_\mathrm{g}}
\newcommand{\tg}{t_\mathrm{g}}

\title[MHD simulations of 
    a tidal disruption in GR]{Magnetohydrodynamical simulations of 
    a  tidal disruption in general relativity}
\author[A. S\k{a}dowski et al.]
       {Aleksander S\k{a}dowski$^{1,2}$,	     
        Emilio Tejeda$^{3,4}$,
        Emanuel Gafton$^4$, Stephan Rosswog$^4$\newauthor and David Abarca$^5$\thanks{E-mail: asadowsk@mit.edu (AS); 
etejeda@astro.unam.mx (ET); 
emanuel.gafton@astro.su.se (EG);
stephan.rosswog@astro.su.se (SR); dabarca@camk.edu.pl (DA)} \\
        $^1$ MIT Kavli Institute for Astrophysics and Space Research
77 Massachusetts Ave, Cambridge, MA 02139, USA\\
$^2$ Einstein Postdoctoral Fellow,\\
$^3$ Instituto de Astronom\'ia, Universidad Nacional
        Aut\'onoma de M\'exico, AP 70-264, Distrito Federal 04510, Mexico\\
  $^4$ Department of Astronomy and 
Oskar Klein Centre, Stockholm University, AlbaNova, SE-10691
Stockholm, Sweden\\
$^5$ Nicolaus Copernicus Astronomical Center, Bartycka 18, Warsaw
00-716, Poland}

\begin{document}

\maketitle

\label{firstpage}

\begin{abstract}

  We perform hydro- and magnetohydrodynamical general relativistic
  simulations of a tidal disruption of a $0.1\msun$ red dwarf
  approaching a $10^5\msun$ non-rotating massive black hole on a
  close (impact parameter $\beta=10$) elliptical (eccentricity
  $e=0.97$) orbit. We track the debris self-interaction,
  circularization, and the accompanying accretion through the black
  hole horizon. We find that the relativistic precession leads to the
  formation of a self-crossing shock. The dissipated kinetic energy heats
  up the incoming debris and efficiently generates a quasi-spherical outflow. The
  self-interaction is modulated because of the feedback exerted by the flow on itself. 
  The debris quickly forms a thick, almost
  marginally bound disc that remains turbulent for many orbital
  periods. Initially, the accretion through the black hole horizon results
  from the self-interaction, while in the later stages it is dominated
  by the debris originally ejected in the shocked region, as it
  gradually falls back towards the hole. The effective viscosity in the debris disc
  stems from the original hydrodynamical turbulence, which dominates
  over the magnetic component. The radiative efficiency is very low 
  because of low energetics of the gas crossing
  the horizon and large optical depth that results in photon trapping.

\end{abstract}


\section{Introduction}
\label{s.introduction}

Occasionally, stars pass by close enough to a massive black hole (BH) so that 
the hole's tidal forces overcome the star's self-gravity and the latter is disrupted.
Depending on the 
orbital parameters, a substantial fraction of the star can become unbound -- 
for instance, in the case of a parabolic encounter, 50\% of the mass is expelled \citep{rees88}.
The bound debris returns to the BH at rates that, initially, can significantly 
exceed the Eddington limit, potentially leading to very luminous transient events.

This fact originally motivated the search for the associated observable events. And indeed, a large and increasing number of candidates for tidal disruption events (TDEs) have been detected. Many of them were discovered in optical and UV sky surveys such as GALEX and CFHTLS \citep{gezari08,gezari09}, PTF \citep{cenko12a,arcavi14}, Pan-STARSS \citep{chornock14}, SDSS \citep{vanwelzen11}, ROTSE \citep{vinko14}, ASASSN \citep{holoien14,holoien15}, others
were detected in soft X-rays (\citealp{komossa99,halpern04,maksym10,maksym14,saxton12,lin15}), while
a few unusually long bursts believed to originate in TDEs have also been discovered in the hard X-rays or $\gamma$-rays
\citep{bloom11,cenko12b,brown15} -- see \cite{komossa15} and \cite{roth15}, and references therein.

For constant mass per energy, $dM/dE= {\rm const}$, Kepler's
third law dictates that the bound matter has to be delivered towards the 
black  hole at a rate of $\dot{M} \propto t^{-5/3}$ \citep{rees88,phinney89},
and many TDE light curves could be modelled with this time dependence.
This is to some degree surprising, because the standard picture requires
the stellar debris to first circularize, before energy can be efficiently extracted.
In standard theory it is also necessary for a sufficiently strong magnetic 
field to build up and provide an effective turbulent viscosity. The 
interplay of these two processes determines the ultimate accretion rate 
through the horizon, and it is at least not obvious that accretion from the
circularized flow will follow the same time dependence as the mass return rate.

Another puzzle is the apparent radiative inefficiency of TDEs. If half of 
a solar mass star is accreted on to a BH, say  in a parabolic disruption of 
a solar-type star, and the accretion converts the rest-mass energy into radiation with the
standard efficiency of  $\sim10\%$,  one could expect the total emitted
energy to be of the order of $10^{53}\,\rm ergs$. However, the
observed TDEs do not normally exceed $10^{51}\,\rm ergs$, and exhibit typical peak luminosities of
$\gtrsim 10^{44}~{\rm erg~s}^{-1}$ in soft x-rays and $\sim 10^{43}~{\rm erg~s}^{-1}$ in the R-band 
\citep[e.g.][]{roth15}. 
There are multiple ways
of explaining this discrepancy, but most likely the accretion that
follows a tidal disruption is radiatively inefficient and in
this way very different from the standard picture of disc accretion
which we believe takes place in active galactic nuclei or galactic
BH binaries.

An interesting suggestion to solve these puzzles was recently made by
\cite{piran+15b}, who proposed that the observed
radiation comes only from dissipation in shocks that form when the
debris returns on a close orbit to the BH, is deflected because of relativistic apsidal
precession, and self-intersects the rest of the incoming stream. In such a way one could
explain the time dependence of the observed light curves -- the
dissipation in shocks is indeed proportional to the debris return
rate. However, a significant fraction of the shocked debris is likely to
remain in the innermost region and one still has to explain why it is
not accreting in a radiatively efficient way. If, however, the self-crossing
takes place relatively far from the BH, then it will
result in changing the orbits of the debris into very eccentric
ellipses \citep{piran+15b}. If debris ends up under the horizon after falling along such
an orbit, it will not extract energy efficiently -- since the specific energy
of such an elongated orbit is close to zero. However, the relativistic
precession in the innermost region may significantly change this
picture.

These are not the only puzzles related to the final fate of stars
disrupted by supermassive BHs. Analytical models are obviously
limited in their applicability and cannot reliably track  the nonlinear stages of debris
evolution, which instead requires numerical simulations.
Such calculations, however, are technically very challenging, since after being disrupted the initial 
star becomes spread out across a huge volume with non-trivial shape
which introduces a very large spread in length and time scales.
Moreover, in the vicinity of the BH general-relativistic effects are important, 
while for the disc formation and evolution radiative transfer and possibly magnetic fields should also be included. So far,
tidal disruptions have been evolved mostly using the smooth particle
hydrodynamics (SPH) method  \citep{laguna93b,kobayashi04a,rosswog08a,rosswog09a,ramirezruiz09,hayasaki13,hayasaki15,bonnerot15,coughlin15} 
or hydrodynamical codes with adaptive mesh refinement \citep{guillochon13,guillochon14}. 
Both methodologies can deal well with adapting to the mass distribution, but to date
none of the existing approaches accounts properly for the relativistic effects close
to the BH, which affect both geodesic motion and gas dynamics.

In this work we follow a hybrid approach. We use an SPH code that accurately accounts for the self-gravity
of the star and its debris to calculate the disruption of a red dwarf ($0.1\,\msun$) 
approaching a supermassive BH ($10^5\,\msun$) on a very close (impact parameter, 
$\beta=10$), elliptical ($e=0.97$) orbit. 
Once the debris starts to return towards the BH, self-gravity is 
no longer important and we translate the output from the SPH simulation 
on to a grid-based GRMHD code where we follow
the subsequent evolution of the debris on its way towards the
BH. A similar approach has been adopted by
\cite{shiokawa+15}. We performed two grid-based simulations, 
one purely hydrodynamical and another one
that accounts for the magnetic field evolution from the onset of the grid-based 
calculation.

The paper is organized as follows. We describe the SPH and the
GRMHD methods in Sections~\ref{s.sph} and \ref{s.grmhd},
respectively. Section~\ref{s.tidal} discusses the tidal disruption we
consider. Results for the non-magnetized evolution are presented in
Section~\ref{s.hydro} and Section~\ref{s.mhd} describes the
magnetized case. In the Discussion (Section~\ref{s.discussion}) we
comment on various caveats that have to be taken into
account. Finally, in Section~\ref{s.summary} we summarize our findings
and the conclusions derived from them. 

\section{Numerical methods}
\label{s.methods}

\subsection{SPH code}
\label{s.sph}

The first stage of the tidal disruption process is simulated with a
smoothed particle hydrodynamics (SPH) \citep{lucy,GM77} code. 
SPH is fully Lagrangian and conserves mass, momentum, angular momentum
and energy by construction. The major advantage in this context is 
that the particles directly trace the matter flow and no computational
effort is wasted on empty regions. Therefore, the spreading of the initially
well-localised spherical star into a huge, geometrically complicated shape
can be followed with virtually no additional computational burden. For Eulerian
methods, even with an adaptive mesh, covering the huge volume while 
resolving the relevant density structures is severely challenging, albeit possible. 
See,  e.g.,~\cite{monaghan05,rosswogsph,springel10,price12a,rosswog15c} 
for recent reviews on SPH, and \cite{rosswog15b} for recent improvements 
of the method.

In this project we use a relativistic extension of the Newtonian SPH code described in
detail in \cite{rosswog08b}. Previous versions of this code have been used for studying the
tidal disruption of white dwarf stars by intermediate-mass BHs
\citep{rosswog09a} and of main-sequence stars by supermassive BHs \citep{gafton15}. The main modification to this code is 
a new formulation of the evolution equations that allows us to
introduce an exact relativistic treatment of the gravitational 
and hydrodynamical accelerations acting on each SPH particle. See \cite{TGR} for further details
and validation of this new implementation. For the present
project, the gravitational acceleration due to the central BH corresponds to 
the exact acceleration acting on a fluid element in Kerr
spacetime as computed in Kerr-Schild coordinates.
The total force 
exerted on each SPH particle is complemented 
by an approximate treatment of the fluid's self-gravity computed 
in a Newtonian fashion using a binary tree as in \cite{benz90}.
Hydrodynamical shocks are captured by means of an
artificial viscosity scheme based on time-dependent parameters that
ensure that it is applied only where and when needed
\citep{morris97,rosswog00}, in addition a switch to suppress 
dissipation in pure shear flows \citep{balsara} is applied.

A red dwarf star is to a good approximation convective
and therefore well-described by a $\Gamma=5/3$ polytropic
equation of state (EOS). We therefore use this EOS 
for the initial star and throughout the whole hydrodynamic evolution, 
both in the SPH and the subsequent general-relativistic magnetohydrodynamics (GRMHD)
simulation. During the hydrodynamic evolution parts of the flow
may become radiation-dominated and therefore their EOS might 
in these regions be better described by a $\Gamma$
closer to 4/3. But nevertheless, for this initial study we stick to an exponent
of $\Gamma=5/3$, and leave more sophisticated approaches to the
EOS for future studies.

\subsection{GRMHD code}
\label{s.grmhd}

\koral was developed in recent years
\citep{sadowski+koral,sadowski+koral2} and is capable of performing
global general relativistic (GR) radiative (R) magnetohydrodynamical (MHD) simulations of
gas dynamics in strong gravitational field of compact objects. \koral uses a shock-capturing
Godunov-based scheme and evolves the equations of GRRMHD in a
conservative form. The code adopts the Lax-Friedrichs Riemann solver
and uses the MINMOD reconstruction with the diffusivity parameter
$\theta=1.5$. Computations are
done in the appropriate fixed general relativistic space-time as
described by the Kerr metric of the spinning BH.
\koral uses Kerr-Schild coordinates which allow
the inner computational grid boundary to be inside the BH horizon.  As
a result, the inner boundary is causally disconnected from the
domain. \koral can be used with
arbitrary coordinates, i.e., the grid points may be arbitrarily 
concentrated in the region of interest. \koral is energy-conserving to machine
precision. This is very useful for numerical studies of accretion
since energy conserving schemes correctly convert any magnetic or
turbulent energy lost at the grid scale into entropy of the gas.  

\koral performs the computations in geometrical units. From now on
we will report length and time in gravitational units,
$\rg=\mathrm{G}M_\mathrm{BH}/\mathrm{c}^2$ (also denoted here
sometimes as $M$) and 
$\tg=\mathrm{G}M_\mathrm{BH}/\mathrm{c}^3$,
respectively. We adopt the following definition
for the Eddington mass accretion rate,
\be
\label{e.medd}
\Medd = \frac{L_{\rm Edd}}{0.057 c^2},
\ee
where $L_{\rm Edd}$ is the 
Eddington luminosity and $0.057$ is the radiative efficiency of a thin disc around 
a non-rotating BH (\citealp{novikov73}). According
to this definition, a standard thin, radiatively efficient disc accreting at
$\Medd$ would have a luminosity of $L_{\rm Edd}$. 
Table~\ref{t.units} gives conversion factors between CGS and
geometrical or Eddington units.

\begin{table}
\centering
\caption{Unit conversion}
\label{t.units}
\begin{tabular}{lr}
\hline
\hline
GU &  CGS \\
\hline
$\rg$ & $1.477\times 10^{10}\,\rm cm$\\
$\tg$ & $0.493\,\rm s$\\
\hline
Edd. &  CGS \\
\hline
$\Ledd$ & $1.25 \times 10^{43}\,\rm erg/s$\\
$\Medd$ & $2.48 \times 10^{23}  \,\rm g/s$\\
\hline
\hline
\end{tabular}
\end{table}

\section{Tidal disruption setup}
\label{s.tidal}

For this work we have followed the tidal disruption of a 
red dwarf initially placed along a bound orbit about a non-rotating 
BH with mass $M_\mathrm{BH}=10^5\,\Msun$. The red dwarf has a mass of $M_\star=0.1\,\Msun$, 
an initial radius of $R_\star=0.15\,\Rsun$, and is modelled using $2\times 10^5$ SPH particles initially arranged to reproduce the density profile corresponding to the solution of the Lane--Emden equations for
a $\Gamma=5/3$ polytrope. During the simulation, the fluid reaction to dynamical compression and expansion is also modelled using a polytropic equation of state with adiabatic exponent $\Gamma=5/3$. After its initial setup, the star is subject to a relaxation procedure so that
the particles find their equilibrium positions and very closely approximate the true hydrostatic equilibrium
solution \citep{rosswog07}. Subsequently, the star
 is placed at a distance of $r_0 = 4\,r_\mathrm{T}$ from
the central BH, 
where $r_\mathrm{T}=\left(M_\mathrm{BH}/M_\star\right)^{1/3}R_\star\simeq71\,\rg$ is the
tidal radius. The star is imparted an initial velocity that places 
its center of mass along a bound orbit with eccentricity $e=0.97$ and impact parameter
$\beta\equiv r_\mathrm{T}/r_\mathrm{p}=10$. This value of the impact parameter corresponds to 
a pericenter distance of $r_\mathrm{p} \simeq 7\,\rg $. The apocenter distance is  
$r_\mathrm{a} \simeq 464\,\rg $.

\begin{figure*}
\includegraphics[width=1.9\columnwidth]{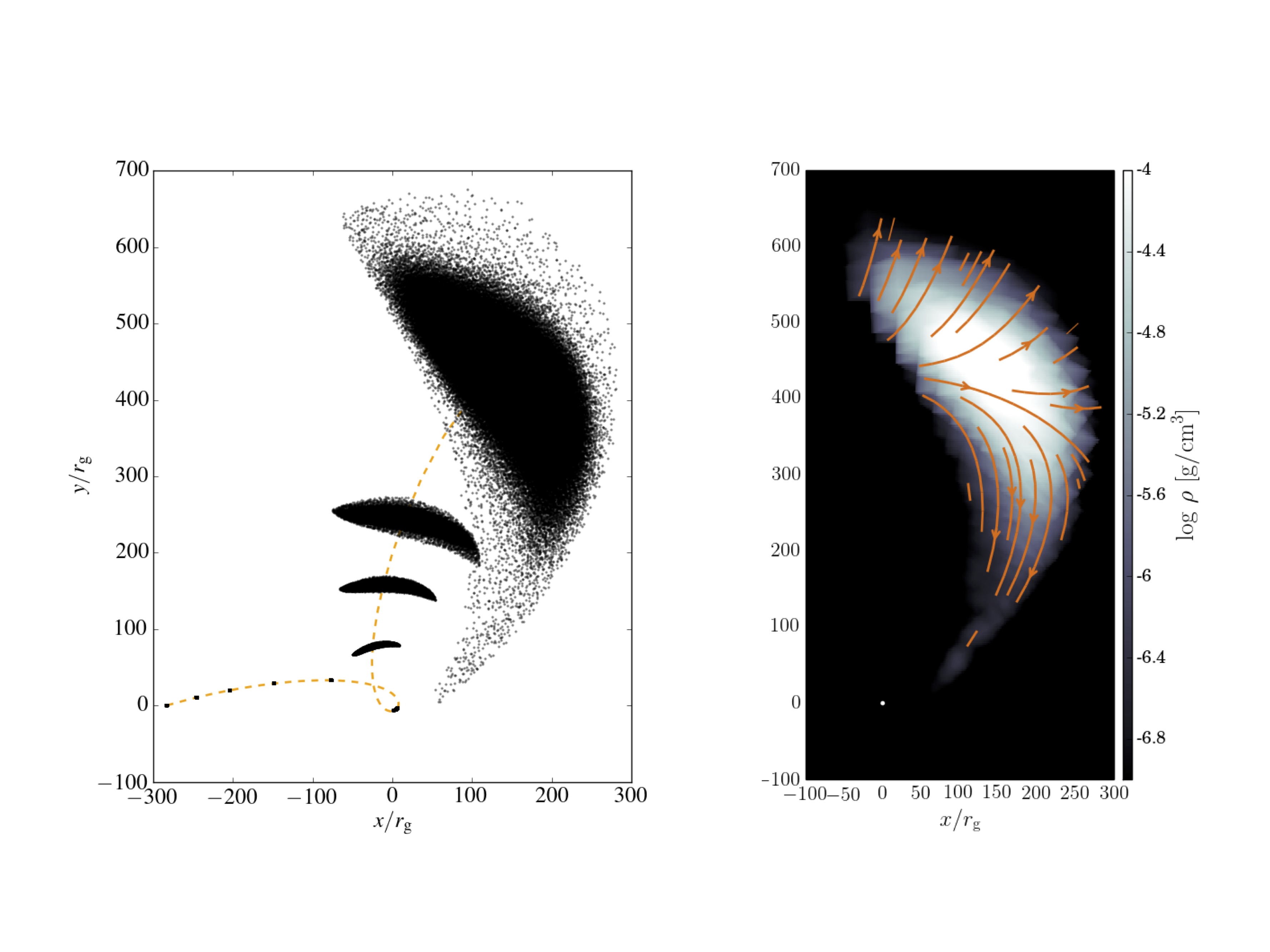}

\vspace*{-1.2cm}

\caption{\textit{Left:} Snapshots of the tidal disruption from the SPH
simulation. The last one corresponds to the initial condition of the
\koral simulation. Density of points corresponds to the surface
density of particles. \textit{Right:} Density on the equatorial plane and
gas velocity at the onset of the \koral simulation, just after
translating the SPH output on the grid. }
\label{f.sph}
\end{figure*}

The initial approach of the star towards the BH and the disruption itself
are followed with the SPH code (see the left panel in Figure~\ref{f.sph} for representative
snapshots). From this figure it is clear that the star is completely disrupted after
the first pericenter passage. After disruption, the whole stellar gas remains energetically bound to 
the BH. Nevertheless, by the end of the SPH simulation, only the most tightly bound material
has started to fall back on to the BH (though none has yet been accreted),
and the less bound material is still expanding outwards.
 This first part of the simulation comprises about 
$8\times10^3\,\mathrm{s}$ of coordinate time and finishes when the head of the disruption debris has
returned to a distance of less than $100\,\rg$ from the BH.

\begin{figure}
\includegraphics[width=1.\columnwidth]{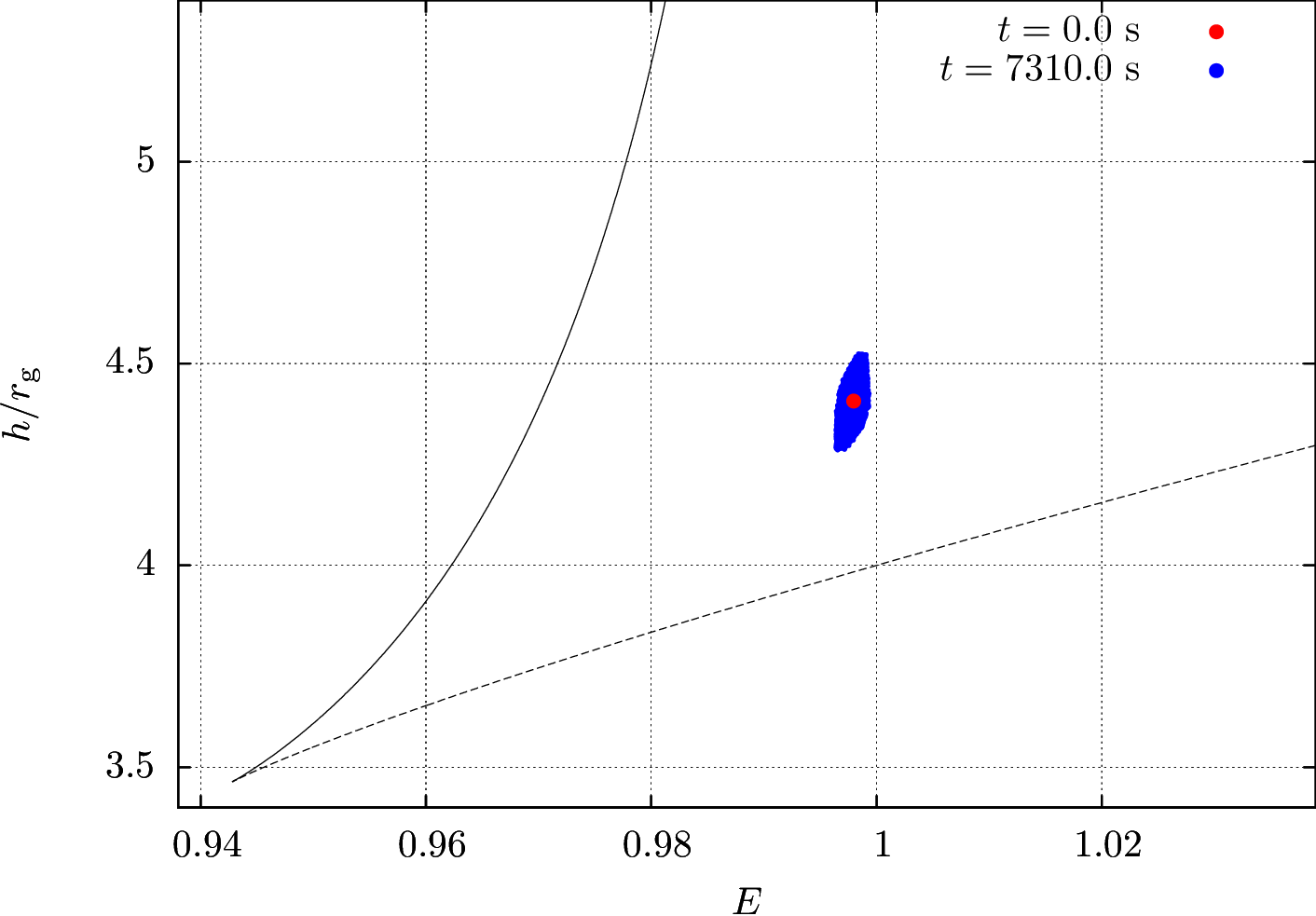}
\caption{Distribution in relativistic specific energy $E$ 
and specific angular momentum $h$ after disruption, at the end of the 
SPH simulation. The red dot indicates the initial values of these 
quantities at the beginning of the simulation. The domain is delimited 
by two lines. The upper branch (solid line) corresponds to stable circular 
motion -- geodesic motion can only be found to the right of this line. 
The lower branch (dashed line) corresponds to unstable circular motion --
any point located below this line is in a plunging trajectory towards the
BH.  }
\label{f.Evsh}
\end{figure}

\begin{figure*}
\includegraphics[width=\linewidth]{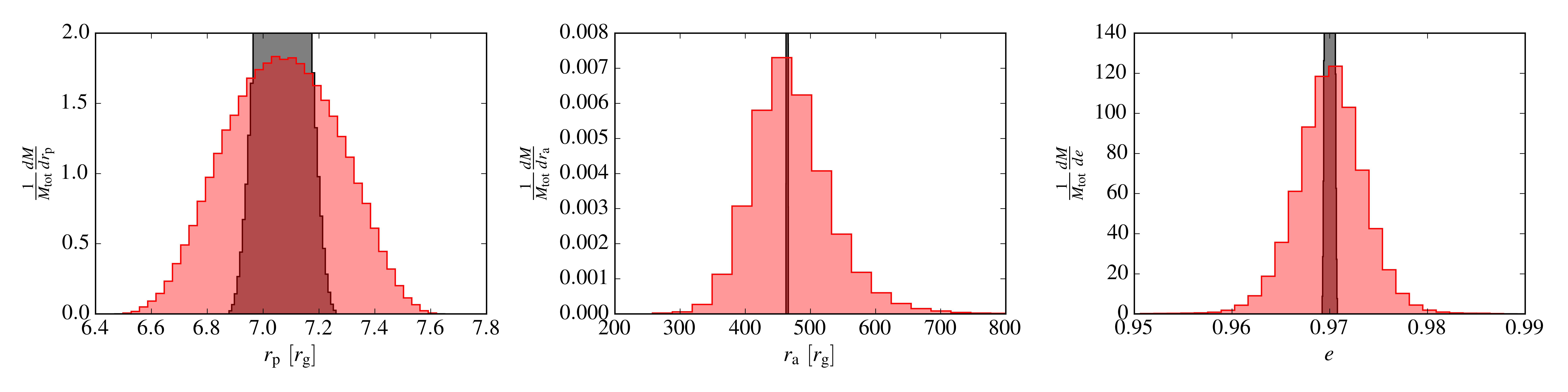}
\caption{Mass distribution of pericenter distances (left panel), apocenter
distances (middle panel) and eccentricities (right panel) before and after disruption (black and red lines,
respectively). The two stages shown in each plot also correspond to the first and the last snapshots of the SPH simulation.}
\label{f.peri}
\end{figure*}

The final distribution of specific energy $E$ and angular momentum $h$ after disruption 
is shown in Figure~\ref{f.Evsh}. In Figure~\ref{f.peri} we show
the mass distribution with respect to pericenter distances, apocenter distances and 
eccentricities before and after disruption. By the end of the disruption process, all these three quantities 
are distributed almost symmetrically about their initial values.

\subsection{Mapping on to the KORAL grid}

In the second phase of this study, we  follow the circularization process 
and the build-up of the accretion disc using the \koral code which is better suited to deal 
with this part of the evolution. To this end, the fluid properties at the end of the SPH
simulation were interpolated on to the Eulerian grid used in \koral.

The \koral coordinates are a modified version of Kerr-Schild
coordinates \citep{sadowski+dynamo}. The grid points are spaced
logarithmically in radius, uniformly in azimuth, and are concentrated towards the equatorial
plane. The adopted resolution was $256\times 128\times 96$ cells in radius, polar
angle and azimuth, respectively. The inner radial boundary was placed
inside the BH horizon, at $r=1.85\,r_{\rm g}$, the outer at
$r=1000\,\rg$. In the polar direction, the grid covered the region between 
$0.0075\,{\rm rad}<\theta<\pi-0.0075\,{\rm rad}$. The angular vertical extent of the
cells at the midplane was $\sim 0.01\,{\rm rad}$. The grid extended over
the full $2\pi$ angle in the azimuth. 

The output of the SPH simulation was translated on to the \koral grid by
interpolating the relevant quantities (density, temperature and velocities) from
the SPH data at the locations corresponding to the centres of the \koral grid
cells. In the end, for consistency, the total amount of gas in the
\koral grid was integrated and normalized in order to recover exactly the mass of
the disrupted star ($0.1M_\odot$). The grid cells with no corresponding SPH particles
were filled with residual atmosphere at rest. The density of that
atmosphere was typically 10 orders of magnitude lower than the maximum
density in the stellar debris.

The last snapshot of the SPH simulation shown in the left panel in
Fig.~\ref{f.sph} reflects the disruption stage when the debris was
translated on the \koral grid. The right panel in the same figure shows
the equatorial plane density distribution at that moment, as
represented within \koral. The streamlines reflect the gas
velocity. Although the outermost parts of the disrupted star are
moving outwards, they are bound and will ultimately fall on to the
BH. From that point on, the disruption was evolved within \koral,
with the same polytropic index as in the SPH simulation,
$\Gamma=5/3$. 

For the \koral simulation we adopted outflow boundary
  conditions at the inner radial edge located inside the BH horizon
  and outflow/no inflow boundary conditions at the outer edge. Both
  the non-magnetized and magnetized simulations were run for roughly
  $130000\,\tg$, corresponding to approximately $18\,\rm h$.

\section{Results for the non-magnetized star}
\label{s.hydro}

We start describing the evolution of the non-magnetized debris. In
Section~\ref{s.mhd} we discuss the changes in this picture when 
a magnetic field is introduced.

\subsection{Self-crossing}
\label{s.self}

Translated on to the \koral grid, the stellar debris continues its fallback towards the
BH.
The pericenter of the original stellar orbit was at
$r=7\rg$, but the disruption of the star led to a significant
spread in orbital energies.  The pericenters of the returning
debris cover a wider range of radii, with the most bound debris 
approaching the BH almost directly, at orbits close to plunging ones.

\begin{figure}
\centering\includegraphics[width=.9\columnwidth]{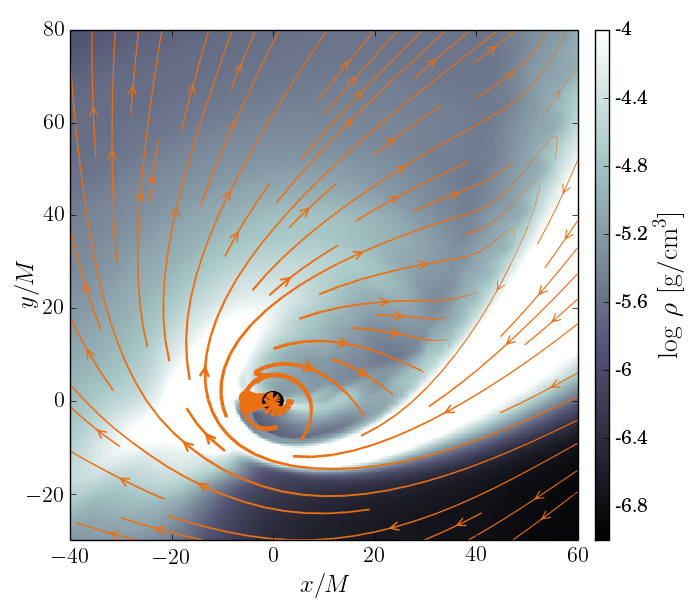}\vspace{-.65cm}
\centering\includegraphics[width=.9\columnwidth]{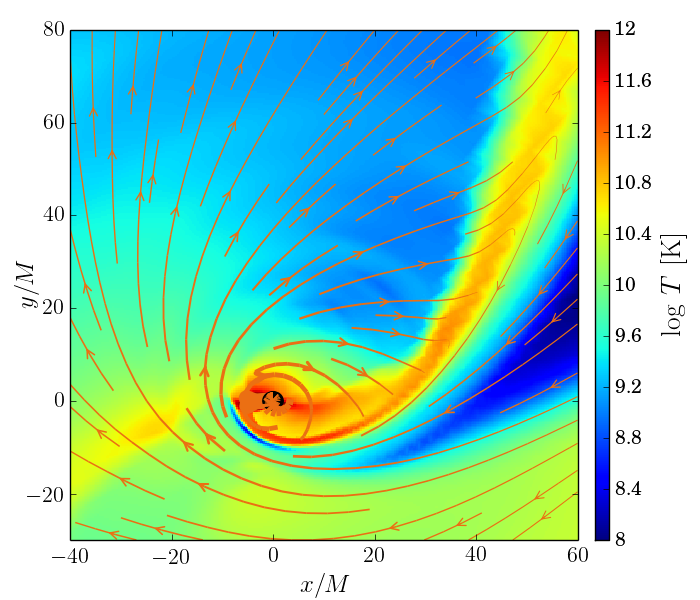}\vspace{-.65cm}
\centering\includegraphics[width=.9\columnwidth]{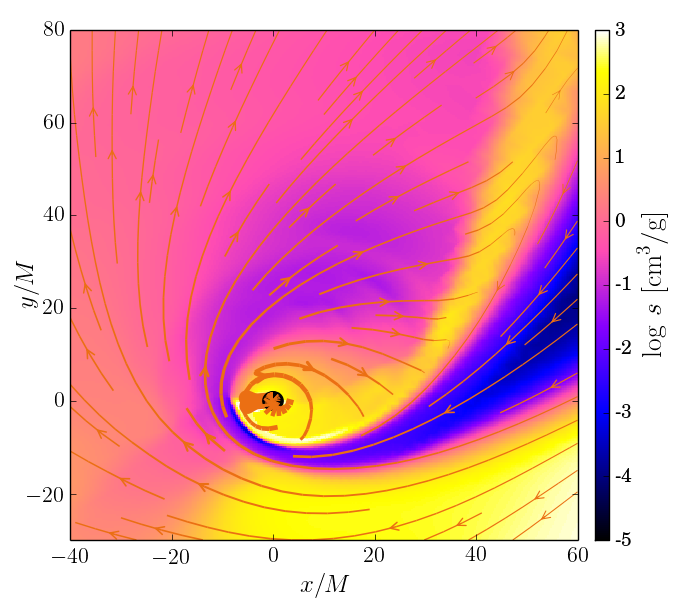}
\caption{Top to bottom: Density, temperature and entropy 
 distributions at $t=10000\,\tg$. Streamlines reflect the
  gas velocity direction and magnitude. }
 \label{f.hd.self}
\end{figure}

The debris is significantly deflected due
to relativistic precession, and does not continue along its
original elliptical orbit. Instead, it hits the incoming stream of gas
relatively close to the BH. Because of the significant width of the
incoming debris stream (a characteristic feature of elliptical
orbits), self-crossing takes place at a wide range
of radii, $10\lesssim r/\rg \lesssim 100$ (see top panel in
Fig.~\ref{f.hd.self}). Effective collision is possible because
there is no BH spin-related precession of the orbital plane of the debris.

\begin{figure}
\centering\includegraphics[width=.8\columnwidth]{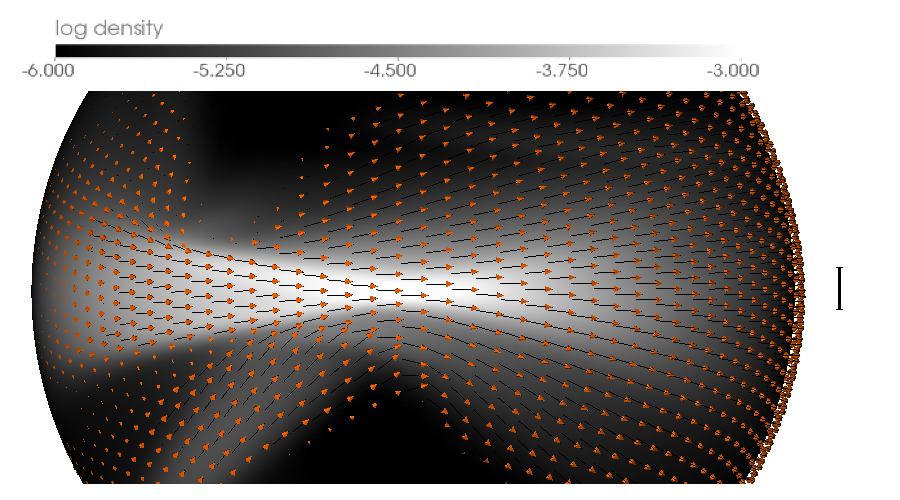}
\caption{Logarithm of gas density on a spherical
    slice taken at radius $r=10\,\rg$. This choice corresponds to the
    smallest crossection of the nozzle, visible in the center. Because
  of the small pericenter radius and eccentricity of the orbit, the
  nozzle is relatively thick. The length of the bar on the
  right corresponds the vertical extent of 10 grid cells. The arrows reflect
the gas velocity.}
\label{f.hd.self.sph}
\end{figure}

For the same reason, i.e., relatively compact stellar debris
that has experienced strong tidal disruption close to the BH, the
nozzle that forms at the pericenters of the orbits of the returning
debris is not as thin as it would be for a  parabolic
encounter. Fig.~\ref{f.hd.self.sph} shows the side view of the
spherical slice taken at $r=10\,\rg$, i.e., through the center of the
nozzle. Its vertical extent in the narrowest point is roughly
$1\,\rg$. At that distance and at the equatorial plane this width
is covered roughly by $5$ polar grid cells. We consider it enough to
capture qualitatively the physics taking place there.

Once the gas passes the nozzle it is relativistically deflected and
hits the incoming stream. The middle and bottom panels in
Fig.~\ref{f.hd.self} show the gas temperature and entropy,
respectively. It is clear that the collision results in a shock -- the kinetic energy
of the deflected stream is being dissipated  and results in rapid
increase of temperature and entropy. This ``self-crossing shock''
extends along the edge of the incoming stream, from the innermost
region out to $r\approx 100\,\rg$. The typical velocity of the
  deflected gas hitting the primary stream from aside is
  ca.~$0.2\,\rm c$. The primary gas has a typical temperature of $10^{8.5}\,\rm
  K$. These parameters result in a Mach number of the order of $20$.

In the innermost region, there is a
trace of a secondary shock within $r=20\,\rg$ occurring ahead of the
self-crossing surface. There is no trace of the ``pancake shock'' in the nozzle, which
was thick enough not to produce any rapid shock-related
heating. 

The high gas temperatures are somewhat artificial. Large densities in
the inner region result in large optical depths and in establishing local thermal
equilibrium between gas and radiation. As in accretion flows accreting
at super-Eddington rates, one may expect that it is radiation that
provides most of the pressure support. The gas will therefore have, in
reality, much lower temperature, with the  total pressure of
radiation and gas being equal to the gas pressure in the
hydrodynamical simulation.

\subsection{Outflow}
\label{s.outflow}

The self-crossing of the stellar debris at small radii results in
shock heating. The deflected stream hits the returning tidal stream
with a large velocity, of the order of the free-fall velocity at the
radius of intersection, $r_{\rm I}$. The magnitude of the specific kinetic energy
available for dissipation, $\epsilon_{\rm kin}\approx \rg/r_{\rm I}\approx 1/30$ (taking
$r_{\rm I}=30\,\rg$ as the representative radius of self-intersection)
is much larger not only than the magnitude of 
the binding energy of the initial stellar orbit, $\epsilon_{\rm bind}=-1/2a\approx 0.002$
(where $a$ is the semi-major axis), but also
than the binding energy of  the most bound stellar debris, $\epsilon_{\rm bind,max}=-1/2a_{\rm min}\approx
0.004$, with $a_{\rm min}$ being the semi-major axis of most bound
debris (compare to Fig.~\ref{f.Evsh} which shows the relativistic specific
energy, $E=1-\epsilon$, of the disrupted star). Therefore, even
if only a small fraction of the deflected gas interacts in the shock and
only a fraction of its  kinetic energy is dissipated,
it is likely that the heated gas will get enough thermal energy to become
unbound. At the same time, the gas that deposits its kinetic energy into
the shock will become more bound. One may also expect that shock-heated gas will expand as a
result of the increased thermal pressure.

\begin{figure}
\centering\includegraphics[width=.9\columnwidth]{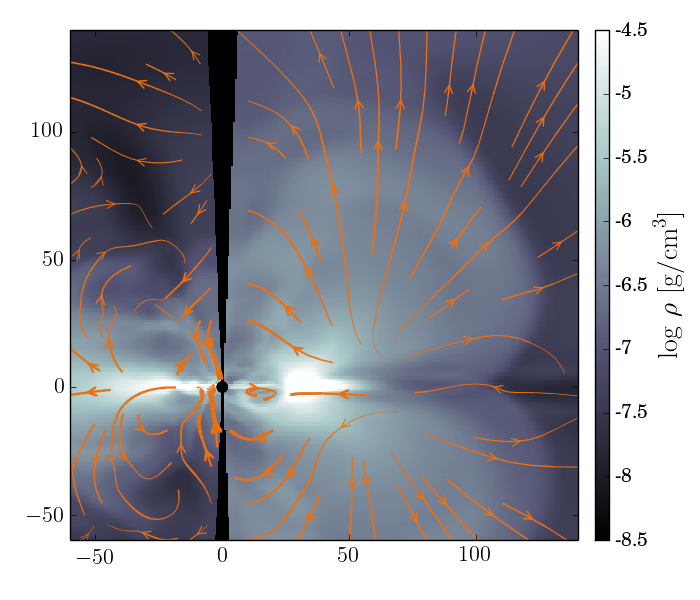}\vspace{-.65cm}
\centering\includegraphics[width=.9\columnwidth]{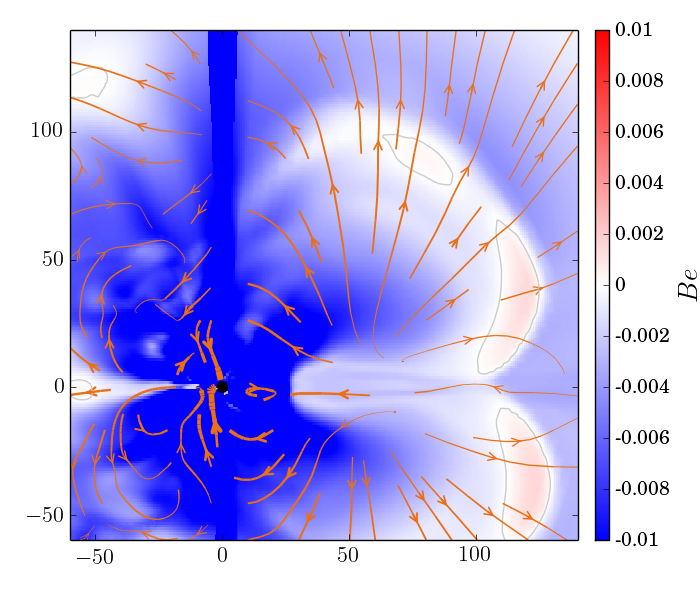}\\
\centering {\vspace{-.3cm}$x/\rg$}
\caption{Distribution of gas density (top) and the corresponding
  Bernoulli function (bottom panel) at the poloidal plane slice taken
  along $y=0$. The
  snapshots correspond to time  $t=10000\,\tg$, the same as in
  Fig.~\ref{f.hd.self}. The grey contour in the bottom panel
  corresponds to $Be=0$.}
\label{f.outflow}
\end{figure}

These two processes are reflected in the panels of Fig.~\ref{f.outflow},
which show vertical slices along $y=0$ of the gas density (top) and the
Bernoulli function (bottom panel) during the self-crossing at an early
stage of the simulation. The Bernoulli function was estimated according
to \citep{sadowski+3d},
\be
Be=\frac{T^t_t-p+\rho u^t}{\rho u^t}.
\label{e.Be}
\ee
Positive and negative values of $Be$ correspond to unbound and bound
gas, respectively. Streamlines in Fig.~\ref{f.outflow} reflect the gas
velocity projected on the poloidal plane.

The right halves of the panels in Fig.~\ref{f.outflow} correspond to the
cross section through the stream of the returning stellar debris. It is
evident that the gas heated in the shock at the edge of that stream
expands and flows out in a quasi-spherical way. The velocity of this
outflowing gas or the thermal energy contained in it may be large
enough to make it unbound. The bottom panel shows the Bernoulli function of gas, with the
original binding energy of the debris, ca.~$-0.003$, 
being rendered in a light blue hue. The outflowing gas is less bound, and some fraction of
it can actually become unbound (red hues). This is possible since,
simultaneously, the debris that loses its kinetic energy in the shock becomes more
bound, as reflected in the deep blue region within $r=30\,\rg$.

\begin{figure}
\includegraphics[width=.95\columnwidth]{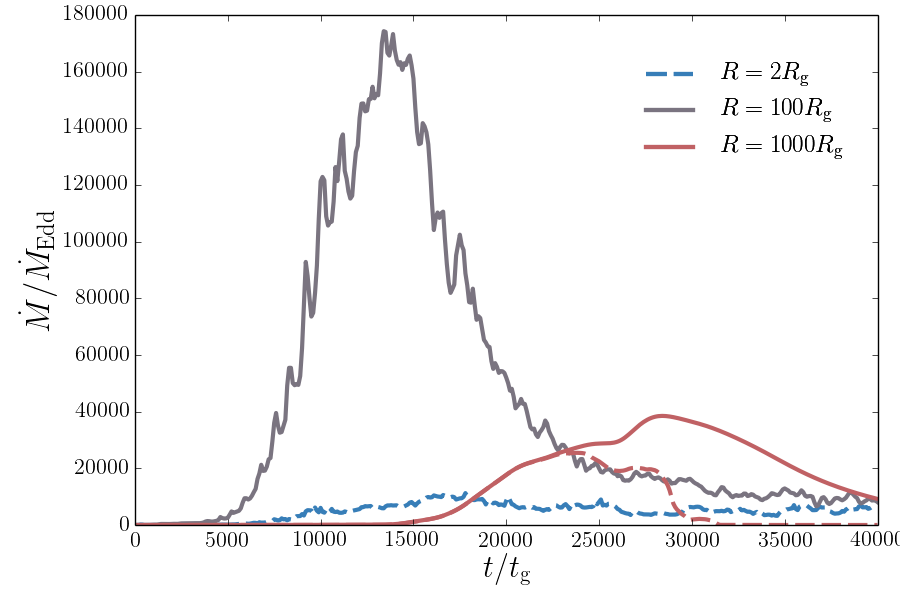}
\caption{Rate (in Eddington units) at which gas flows out through spheres of given
  radii. Solid lines reflect the flux of outflowing mass at $r=100\,\rg$ (grey)
  and $r=1000\,\rg$ (red line). The blue
  dashed line reflects the accretion rate (inward flux) through the BH
  horizon. The red dashed line shows the flux of unbound gas crossing $r=1000\,\rg$.} 
\label{f.inout}
\end{figure}

To quantify the amount of gas outflowing due to shock heating in
the self-crossing region, we calculated the mass flux crossing spheres
of radii $r=100\,\rg$ and $r=1000\,\rg$ (the latter being the outer edge of the computational domain).
Outflow rates calculated this way are shown in Fig.~\ref{f.inout} 
with grey and red lines, respectively. The outflow through the
smaller sphere starts early, at $t\approx5000\,\tg$, when the first debris that
has passed its pericenter reaches a radius $r=100\,\rg$. Most of this mass
flux comes from debris returning to apocentre along its regular
orbit in the equatorial plane. 
 
There is also, however, a significant
component originating from the quasi-spherical outflow caused by the shock
heating of the gas. This debris, being less bound than it was initially,
 can get away from the BH to larger apocentres than those of its original
orbits, and parts of it may even leave the computational domain, a fact that would
not be possible for the original, completely bound debris. The rate
at which gas flows out of the computational domain (at $r=1000\,\rg$) is
denoted by the red line  in Fig.~\ref{f.inout}. The corresponding
dashed line reflects the flux of unbound gas crossing the domain boundary.

Gas starts crossing
that surface only at $t\approx 15000\,\tg$. The lag between the outflow
through the outer and inner spheres reflects the typical velocity of
the outflow in this region, $v\approx 0.1\,\rm c$. The rate at which
the gas flows out through the edge of the computational box is
significantly lower, because most of the gas crossing
the $r=100\,\rg$ surface is bound and thus remains close to the BH. However,
roughly a third of that gas possesses large enough kinetic energies as to reach
$r=1000\,\rg$ and leave the domain, though only a fraction of that gas,
about $40\%$ or $0.015\msun$, is energetic enough to reach infinity.
This means that the bound fraction could in principle fall back on to the volume
that is covered by our computational domain. For the time being, we do not
account for this effect.

\subsection{Periodicity}
\label{s.periodicity}

The rate of gas flowing out through the inner sphere at $r=100\,\rg$
(denoted by the grey line in Fig.~\ref{f.inout}) shows a periodic
pattern on top of the long-term trend. This feature is clearly visible 
especially at early times, for $t<12000\,\tg$. The characteristic period
is $P\approx 800\,\tg$. The magnitude is not strong enough to overwhelm
the general trend reflecting the rate at which mass returns to the BH, but
is clearly imprinted in the outflow rate.

This periodicity comes from the feedback of the deflected debris on
the original stream. The pericenters of the returning debris' orbits are
located very close to the BH, implying significant relativistic
precession. As a result, the self-crossing takes place relatively close
to the BH (at $10 \lesssim r/\rg \lesssim 100$), with the deflected debris
carrying a lot of angular momentum. 
Consequently, the innermost part of the original stream is not only shock heated, but also
gains angular momentum and is therefore pushed at larger radii than its original trajectory. 
Since a larger pericenter radius is accompanied by less 
relativistic precession, the subsequent self-crossing episode takes place at larger
radii than before, which decreases the efficiency of momentum
transfer and allows the incoming stream to revert to (or close
to) its original orbit, i.e., to smaller pericenter radii, resulting once again in more
effective relativistic precession and angular momentum transfer. The cycle then repeats.

The feedback of the deflected stream on the returning one causes a
periodic behavior with the characteristic timescale related to the
time it takes for the gas to come back to the self-interaction region,
i.e., of the order of the Keplerian period at that radius. In the
case of our simulation this period was $t\approx 800\,\tg$, which
corresponds to the Keplerian orbital period\footnote{The frequency corresponding to the
Keplerian period at radius $r$ is in general given by,
\be
\nu_{\rm Kepl}= 1.022\times 10^{-3}
\left(\frac{10\,\rg}{r}\right)^{3/2}\left(\frac{10^6\msun}{M_{\rm
      BH}}\right)\,\rm Hz.
\label{eq.nukepl}
\ee} at $r=25\,\rg$, located, indeed, within
the self-interaction region. 

The periodic pattern is clearly imprinted in the outflow rate measured
at $r=100\,\rg$, but is not visible in the rate of gas leaving the
computational box (measured at $r=1000\,\rg$). The reason is that the
outflow rate is modulated not only through density but also through
velocity. Shells of gas expanding with different velocities quickly
overlap each other and cancel out the periodic pattern. If the periodicity 
were to be detectable in the radiation coming out of the system, photons ought to
separate from the gas relatively close to the BH, at $r\lesssim
100\,\rg$. This is not the case in our simulation. As a result of the short
duration of the disruption, the gas density is large enough to place
the photosphere effectively at the edge of the computational
box. The implications of the feedback
loop described in this Section for more optically thin tidal
encounters will be discussed in detail in
Section~\ref{s.discussion}.

\subsection{Circularization}
\label{s.circularization}

The self-interaction of the stream dominates the evolution of the
system as long as new debris falls into the inner region. Because of
the low eccentricity of the considered stellar orbit ($e=0.97$), all the
stellar debris is bound and ultimately returns to the BH, with the
least bound stellar debris entering the inner region at about 
$t=20000\,\tg$. Starting at that time, both the self-crossing and the 
related ejection of gas cease, while the debris left in the inner region 
continues to interact; this reduces its average eccentricity and drives it
towards circularization.

Figure~\ref{f.hd.tde} shows four snapshots covering the whole
duration of the hydrodynamical simulation. The top panels show the
distribution of gas density in the equatorial plane, while the bottom
panels show the same quantity but on the vertical cross-section along
$y=0$ (as shown with dashed lines in the top panels). The history of the
accretion rate at the BH (discussed in Section~\ref{s.fallback}) is rendered as a line plot
inside the top panels. The snapshots correspond, from left to right, to $t=5000$,
$20000$, $50000$ and $90000\,\tg$.

The leftmost panel reflects the early stage of the self-interaction
phase, when only the tip of the stellar debris has passed its
pericenter. The bottom panel shows only thin structures - the
cross sections through the returning and deflected streams. The next set
of panels correspond to the time when the bulk of the debris enters
the inner region. Significant amount of gas has already accumulated in
the inner region -- it is the gas that became strongly bound after
its pericentre passage and the subsequent interaction with the incoming 
stream. The distribution of this gas, however, is very irregular and
far from circularized. It resembles the structure seen in the late
stages of the simulations performed by \cite{shiokawa+15}, who
simulated a parabolic disruption (with self-crossing occurring 
at a radius $r\approx100\,\rg$) with debris continuosly
flowing in.

The last two sets of panels show late stages of the gas evolution. It
is evident that, by that time, the gas has circularized and no longer
evolves on a dynamical time scale. The debris has settled down in a very thick
disc extending virtually to the edge of the domain. However, most of the
debris is located around radius $r=300\,\rg$, i.e., roughly at the distance
corresponding to the semi-major axis of the stellar orbit. 

\subsubsection{Turbulence}

Somewhat surprisingly, having in mind that we are dealing so far with
a purely hydrodynamical simulation, we see that the debris remains
turbulent even long after the circularization. The turbulence is
inherited from the initial violent stage of self-interaction and the
related feedback loop (Section~\ref{s.periodicity}), which not only
introduces non-axisymmetrical inhomogeneities, but also
breaks the equatorial plane asymmetry (in our case originating in the 
not-perfectly symmetrical SPH initial conditions).

\begin{figure*}
\includegraphics[width=.2505\textwidth]{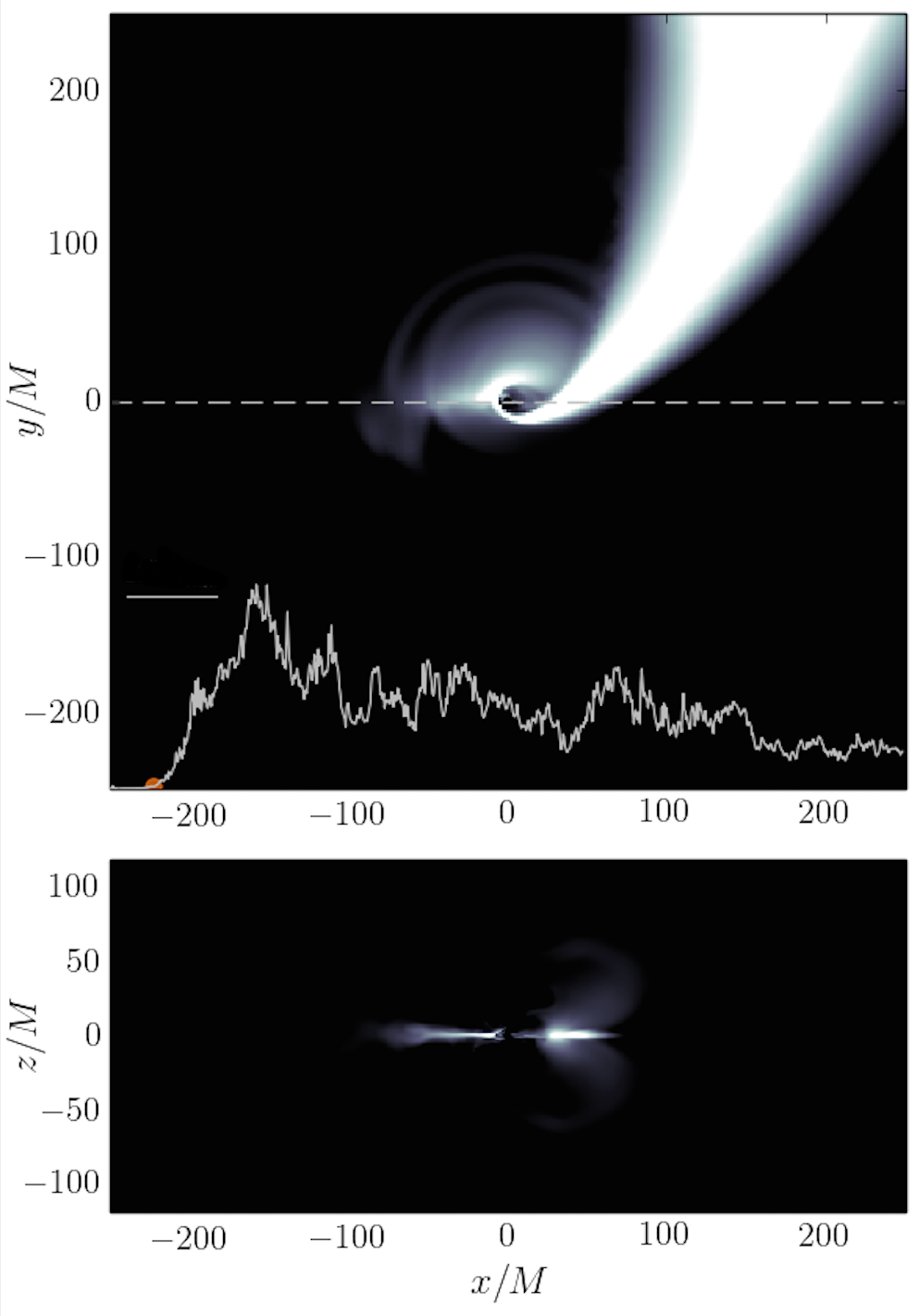}\hspace{-.0cm}
\includegraphics[width=.225\textwidth]{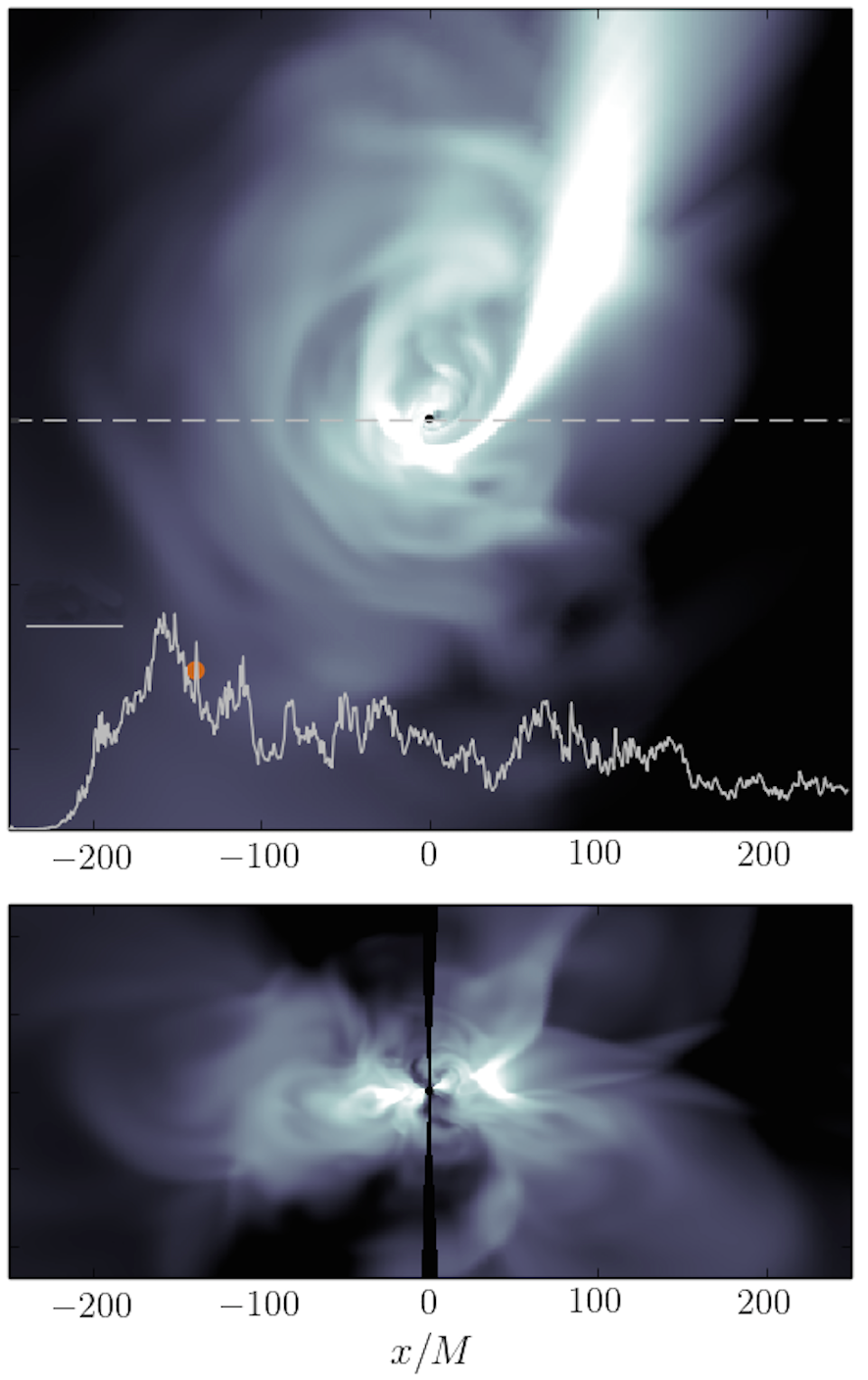}\hspace{-.0cm}
\includegraphics[width=.2245\textwidth]{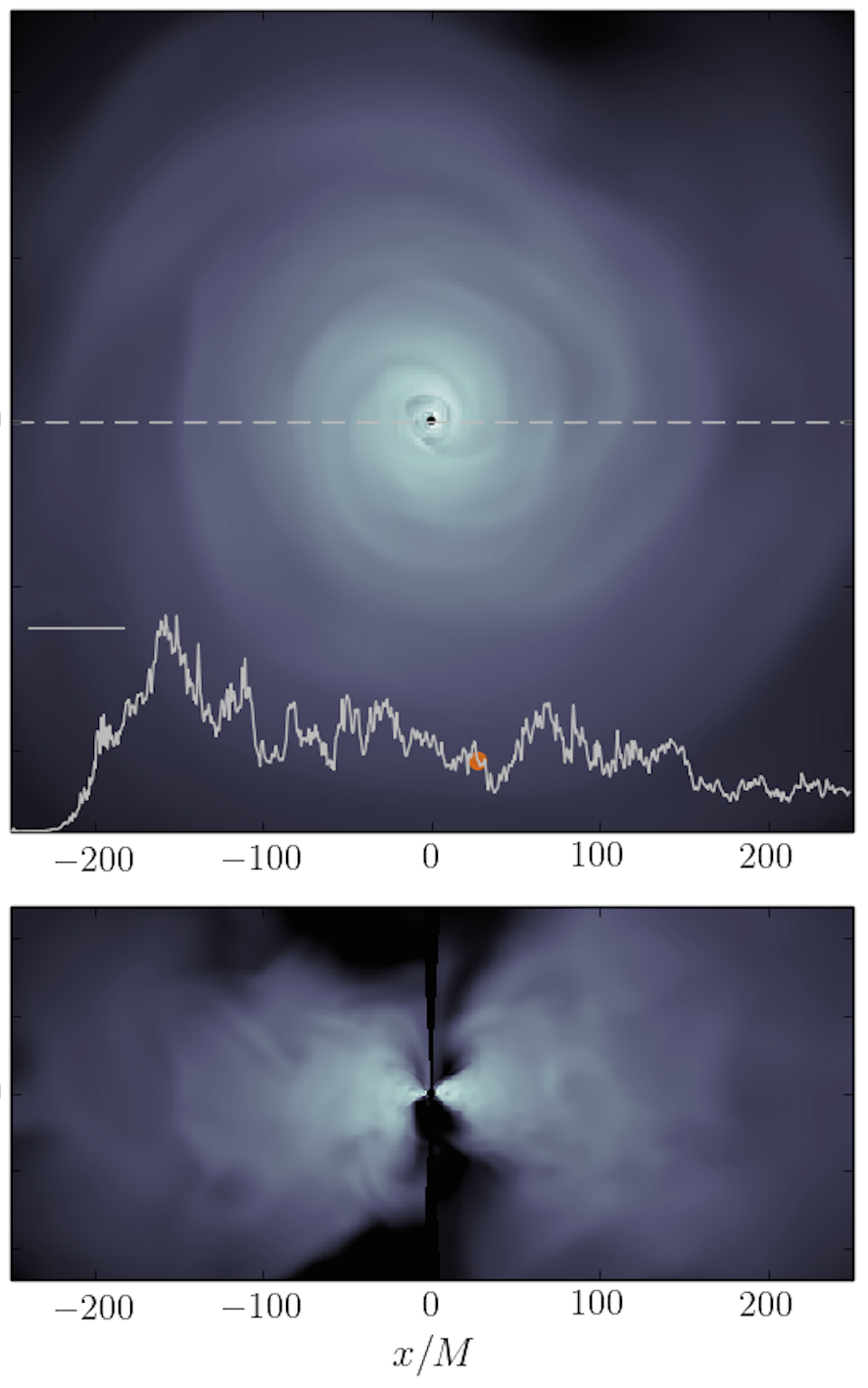}\hspace{-.0cm}
\includegraphics[width=.275\textwidth]{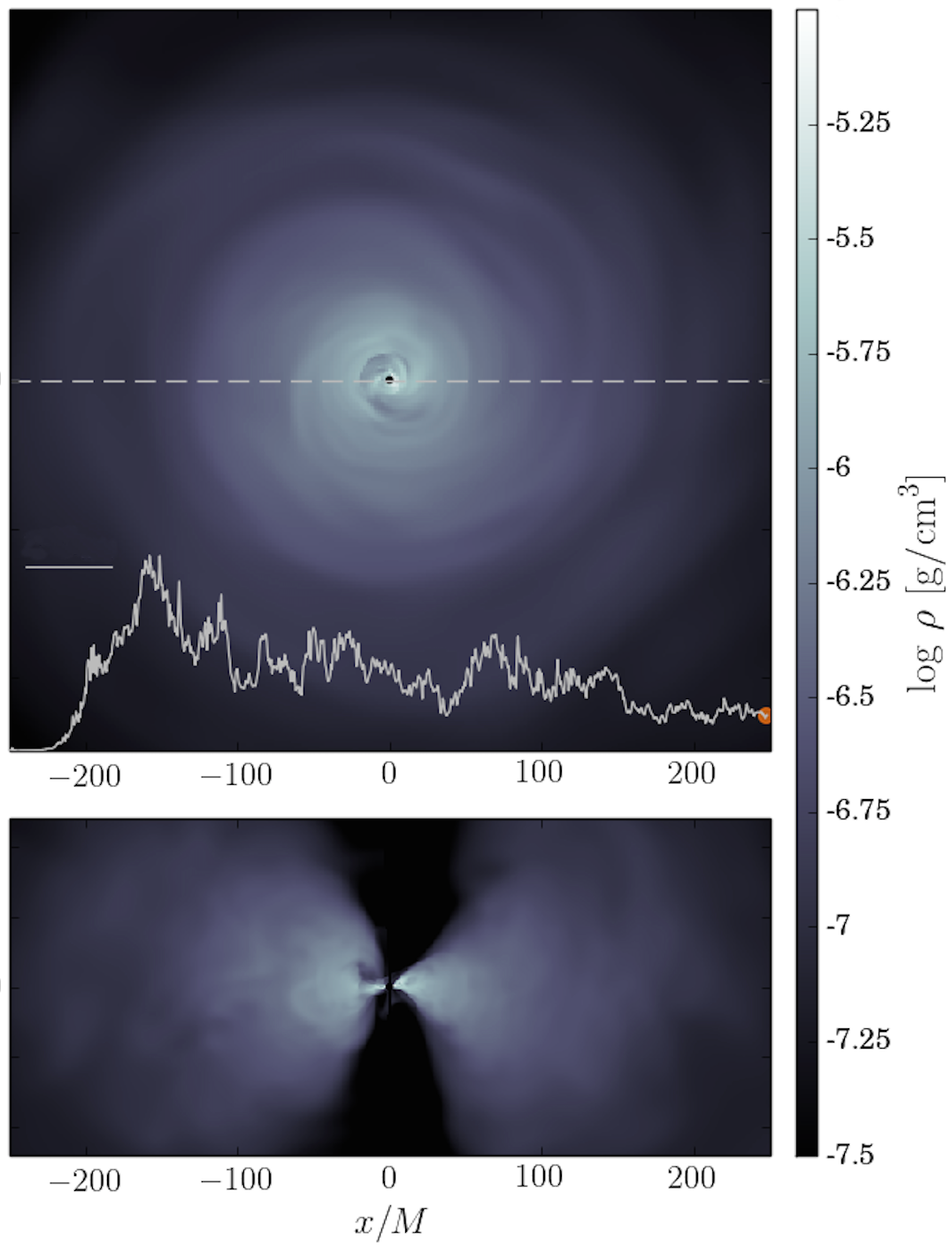}\hspace{-.0cm}
\caption{Snapshots of the equatorial plane (top panels) and a slice
  through the poloidal plane (bottom panels) of the simulated tidal
  disruption event. Colors reflect the logarithm of gas density. The
  history of the accretion rate through the horizon is shown in the
  top panels, with the current time of the snapshot being denoted by a red dot. 
  The horizontal lines corresponds to $10^4\dot M_{\rm
    Edd}$. The snapshots were taken at (left to right) $t=5000\,\tg$,
  $20000\,\tg$, $50000\,\tg$, and $89600\,\tg$ after the start of the simulation.}
\label{f.hd.tde}
\end{figure*}

One might expect
that once the self-interaction terminates, the turbulence would 
decay on  the eddie
turnover time (essentially, the orbital time). However, even after
$130000\,\tg$ from the time when the last debris came into the inner
region, i.e., after ca.~$55$ orbital periods at radius $r=50\,\rg$, the
turbulence still persists (although at a slightly lower level than
initially, compare the third and fourth panels in
Fig.~\ref{f.hd.tde}). 

The question thus arises what process maintains the turbulent
properties of the flow. There are two possibilities, either the flow is
intrinsically unstable or it is perturbed externally.

Differentially rotating tori in equilibrium are known to be unstable
against the \cite{pp-84} instability which requires reflection of
acoustic modes at the boundaries. However, the circularized debris
extends over a large range of radii, and its inner edge is not well
distinguished (see Fig.~\ref{f.rhomdot}). Moreover, the instability is
known to generate $m=1$ modes (\textit{planets}) which are not present
in our simulations. We therefore suspect that the Papaloizou-Pringle
instability is not operating.

\begin{figure}
\includegraphics[width=1\columnwidth]{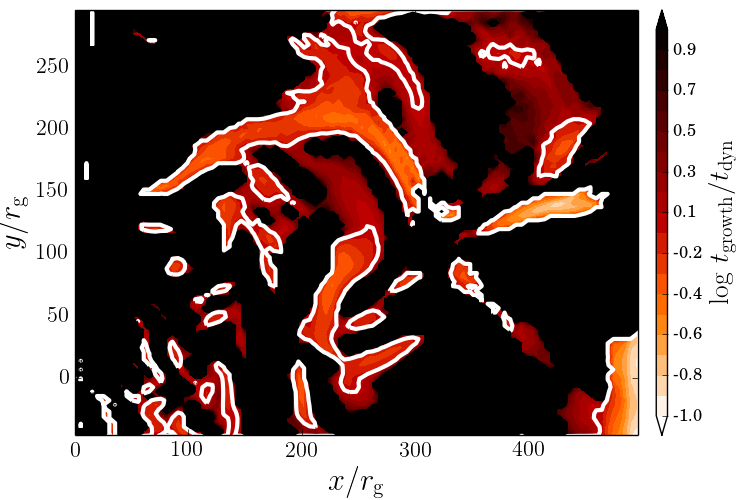}\vspace{-.1cm}
\caption{Convective stability of the hydrodynamical
  simulation calculated using a snapshot data taken at $t=40000\,\tg$. Black regions denote stable regions. Colors
  show the logarithm of the growth timescale over the dynamical
  time. White contours denote regions where the growth timescale is
  shorter than the dynamical time.}
\label{f.hd.conv}
\end{figure}

Another possibility is convection. We studied the convective stability
of the circularized debris following
\cite{tassoul-book}. Fig.~\ref{f.hd.conv} shows the results of the
stability analysis for a snapshot data taken at the onset of the
circularization phase. Black areas reflect the stable regions,
while colors denote unstable ones. The brighter the color, the shorter
the growth timescale of the convective instability. White contours
limit the regions where it is shorter than the dynamical
timescale. The analysis based on the snapshot data shows that
some fraction of the volume is indeed unstable against convection
and that it is likely to grow because of relatively short growth
timescales. We conclude that the debris is partially unstable
against convection and that it may be a crucial factor in keeping the
originally disturbed debris turbulent over a long time.

Another possibility is the external perturbation. Indeed, the debris disc
is constantly affected by the debris (originally ejected during the self-interaction) falling back in
a non-uniform way. It affects the disc by momentum
exchange but may also induce a sonic instability \citep[e.g.,][]{glatzel-88} at the
supersonic shear layer that forms at the border of the disc and the
polar fallback region.

 Singling out which factor dominates in sustaining the turbulence, either
convection or fallback, would require additional studies which we
postpone to the future.

\begin{figure}
\includegraphics[width=.95\columnwidth]{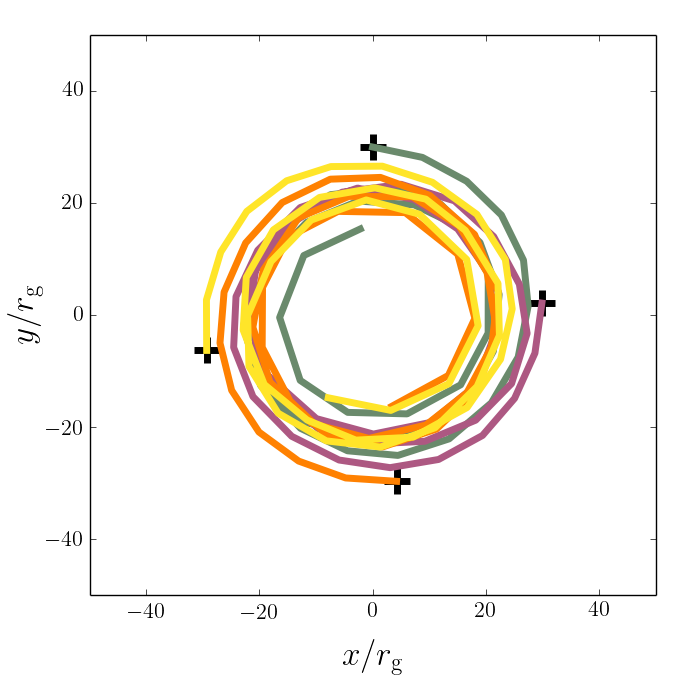}\vspace{-.9cm}
\includegraphics[width=.95\columnwidth]{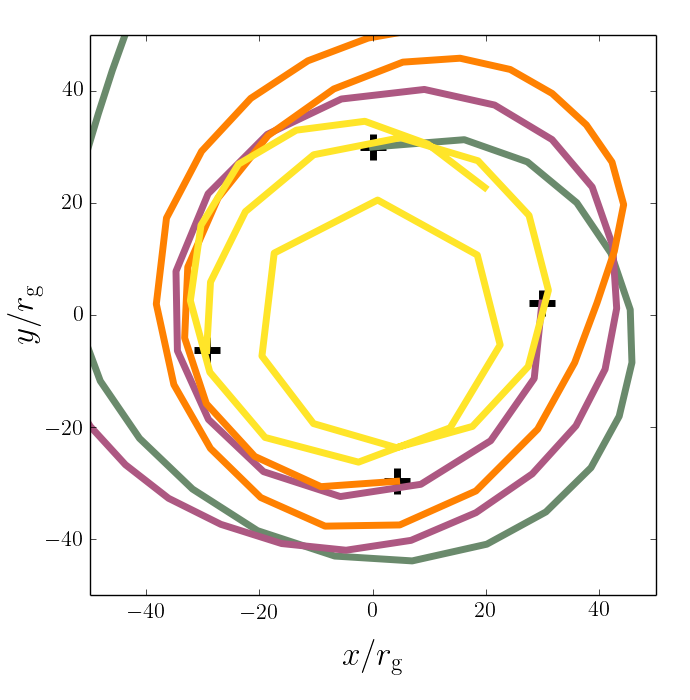}
\caption{Exemplary trajectories of gas parcels in a standard,
  super-critical accretion disc \citep[top,][]{sadowski+3d} and in the simulated TDE
  circularized debris (bottom panel). The trajectories start at the
  same moment of time ($t=10000\,\tg$ for the super-critical disc and
  $t=50000\,\tg$ for the TDE debris), in the equatorial plane at the
  locations denoted by black crosses. }
\label{f.trajs}
\end{figure}

The flow circularizes forming
a well defined toroidal rotating structure, though the gas parcels do
not move along perfectly circular orbits, but rather violently move in and out with
radial velocities of similar magnitude to the orbital velocity. A visual comparison of the dynamics of the circularized debris
and the standard (evolved) accretion disc is given in
Fig.~\ref{f.trajs}, which shows trajectories of four gas parcels that
at the same moment of time were located at $r=30\,\rg$. In the case of 
standard accretion (top panel), gas is only slightly sub-Keplerian and
thus follows almost circular trajectories, gradually approaching the BH. In the case of our circularized
debris (bottom panel), the dynamics is completely different. The low
angular momentum and significant energy of the gas make its
trajectories less uniform, with the radial component of
velocity often comparable to the azimuthal one.

Similar behavior of tidal disruption debris interacting with itself
and circularizing to a turbulent thick disc in which gas parcels do not
  follow circular orbits for a long time was recently observed in SPH
  simulations of interacting streams (Gafton et al.~2016).

  The azimuthal stress resulting from hydrodynamical turbulence is
    significant. The corresponding $\alpha$ parameter is ca.~$0.1$
    near $r=20\,\rg$)
    (Section~\ref{s.viscosity}). The viscosity transports the angular
    momentum outwards and allows the gas deep in the disc to 
    approach the BH. However, as will be discussed in Section~\ref{s.fallback},
    this process is not the dominant source of the gas crossing the BH
    horizon.

Finally, we stress that the observed high power in turbulence results from the
fact that the self-interaction took place close to the BH, what made it
efficient in dissipating and transferring significant amounts of energy.

\subsubsection{Eccentricities}

We will now discuss the circularization process in terms of the
eccentricities of the gas orbits throughout the grid-based simulation. At the
onset of the simulation the gas inherited
eccentricities and orbit orientations from the SPH simulation
(Figs.~\ref{f.Evsh} \& \ref{f.peri}). As a result, the gas approached
the BH along these orbits and only after the pericenter passage did the
orbital parameters start to change.

\begin{figure}
\includegraphics[width=.95\columnwidth]{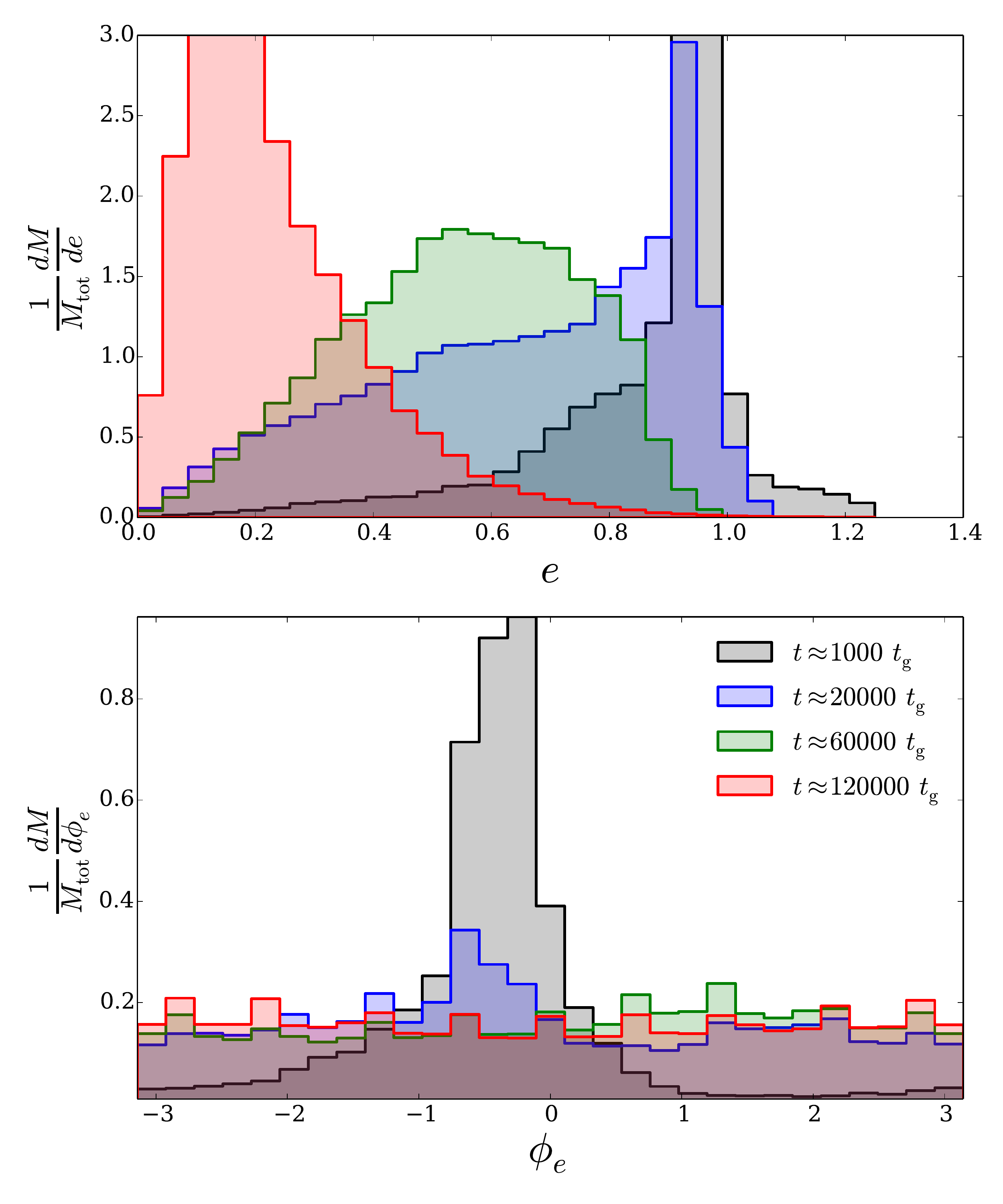}
\caption{Top: Histograms of orbital eccentricity for the gas
at radii between $20\, r_\mathrm{g}$ and $100\,r_\mathrm{g}.$ The
histograms are normalized so that each bin shows the fractional 
mass at that eccentricity. Plots are given for four different times
where each plot corresponds to a time average spanning approximately
$2000\, t_\mathrm{g}$ around that time. In physical units, the times
correspond to 0.56, 11.27, 33.81 and 67.61 hours.
Bottom: Same as the top, except we are plotting the $\phi$ component of the eccentricity vector, so that we 
can track its orientation in the equatorial plane.}
\label{f.eccvst}
\end{figure}

Figure~\ref{f.eccvst} shows how the eccentricities of gas parcels (top
panel) and
the orientations of their orbits (defined through the azimuth of the
eccentricity vector projected on the equatorial plane, bottom panel) evolve with time for
gas within $\pi/6\,\rm rad$ from the equatorial plane. The
eccentricities were obtained by taking the local gas velocity
vector and the local gradient of pressure, and calculating the
effective velocity vector which produces the same net centrifugal
force acting on the gas in the simulation. 

The four histograms shown in the panels correspond to a very early
stage of the
simulation ($t\approx 1000\,\tg$, grey), the end of the self-crossing
stage ($t\approx 20000\,\tg$, blue), and two phases of
quasi-circularization ($t\approx 60000\,\tg$, green, and $t\approx
120000\,\tg$, red). The early distribution of eccentricities closely 
resembles the parameters of the stellar orbit, i.e., the eccentricities
peaking at $e=0.97$ and common orientation for the gas orbits. The
self-interaction initiates the process of circularization, extending the 
distribution of eccentricities to much lower values, and broadening the distribution of
orbit orientations. Once the self-interaction ends, the gas has
already formed  the quasi-disc (see Fig.~\ref{f.hd.tde}), and the initial
orientation of the stellar orbit has been forgotten (see blue and red
histograms). The eccentricities, however, do not become zero. As the
top panel shows, even at the very end of the simulation, the
eccentricities of gas parcels cover a wide range, $0\lesssim e
\lesssim 0.7$. This fact reflects the turbulent nature of the
circularized disc -- on average the gas moves around the BH, but its
motion is far from laminar (Fig.~\ref{f.trajs}).

\subsubsection{Average structure}

\begin{figure*}
\includegraphics[width=.7\textwidth]{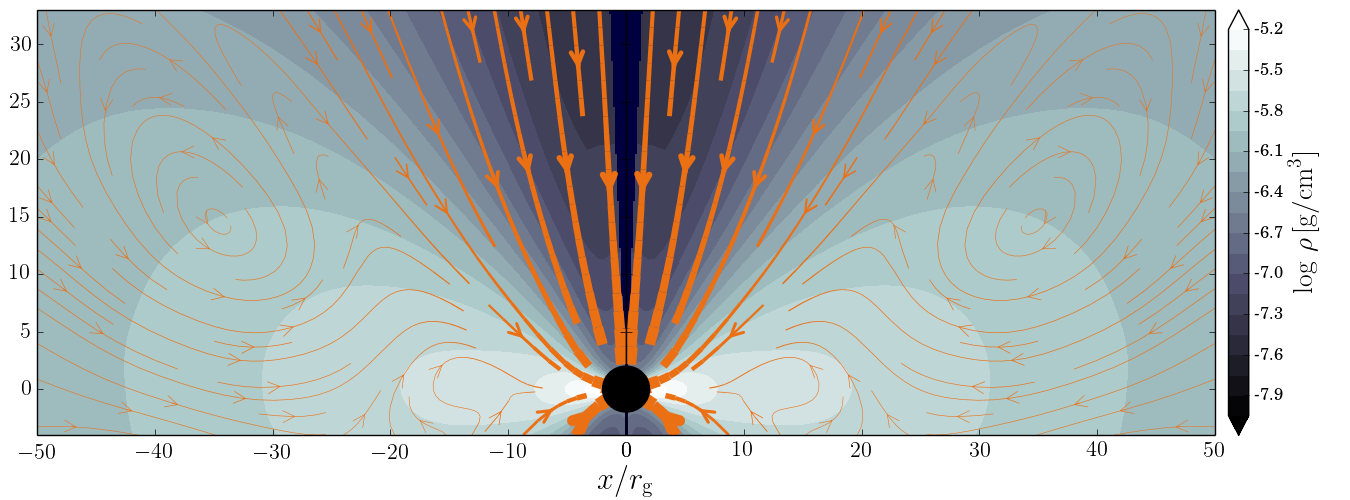}\vspace{-.4cm}
\includegraphics[width=.7\textwidth]{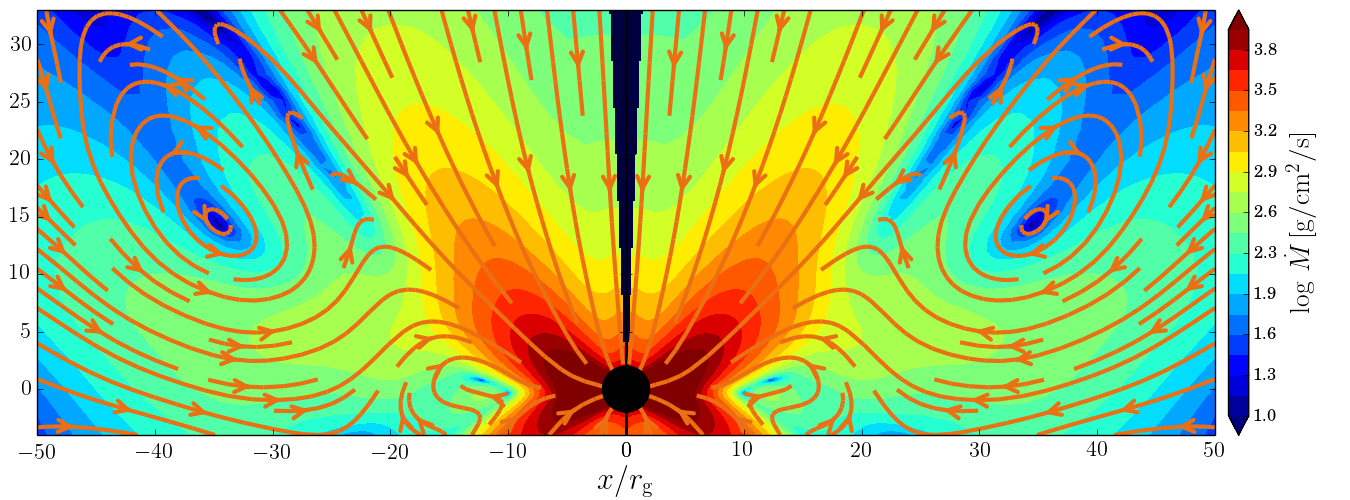}\hspace{-.0cm}
\caption{\textit{Top panel:} Time and azimuth averaged distribution of
  density on the poloidal plane corresponding to the debris evolution
  after the circularization. The streamlines show the poloidal
  component of gas velocity. Their thickness reflects the velocity
  magnitude. \textit{Bottom panel:} Local accretion rate in the
  poloidal plane, $\dot M=\rho
  u^p$, corresponding to the same data. Colors reflect the logarithm
  of the accretion rate.}
\label{f.rhomdot}
\end{figure*}

The average structure of the inner parts of the disc after the circularization is shown
in Fig.~\ref{f.rhomdot}. The top panel shows the distribution of
density in the poloidal plane obtained by averaging in azimuth and in
time over $50000<t/\tg<85000$. Colors reflect the density. The disc
is indeed very thick, with the average density scale-height of
$h/r\approx 2$, larger than the typical $h/r\approx 0.3$ of standard
optically-thick accretion discs \cite[e.g.,][]{sadowski+dynamo}. 

The
streamlines in the same panel reflect the average poloidal components
of velocity. The thickness of the lines is roughly proportional to the
gas velocity. The fastest inflow occurs in the polar region.
Deep inside the disc, the gas hardly moves inwards, with
the average, density-weighted velocity of the order of $10^{-3}$ at
$r=50\,\rg$. Moreover, the velocity pattern is not uniformly pointing
inwards, hinting that the averaging period (ca.~20
orbital periods at $r=40\,\rg$) was not
enough to average out the turbulence. Both facts suggest that the
accretion through the disc is not efficient.

The bottom panel in Fig.~\ref{f.rhomdot} shows the local rate of
accretion ($\dot M=\rho u^p$) in the poloidal plane. It is evident
that the densest interior region of the disc does not contribute
significantly to the rate at which gas reaches the BH
horizon. This quantity is dominated instead by gas falling down along
the edges of the disc. This fallback phenomenon is discussed in
the following Section.

\subsubsection{Specific energy and angular momentum}

\begin{figure}
\includegraphics[width=.95\columnwidth]{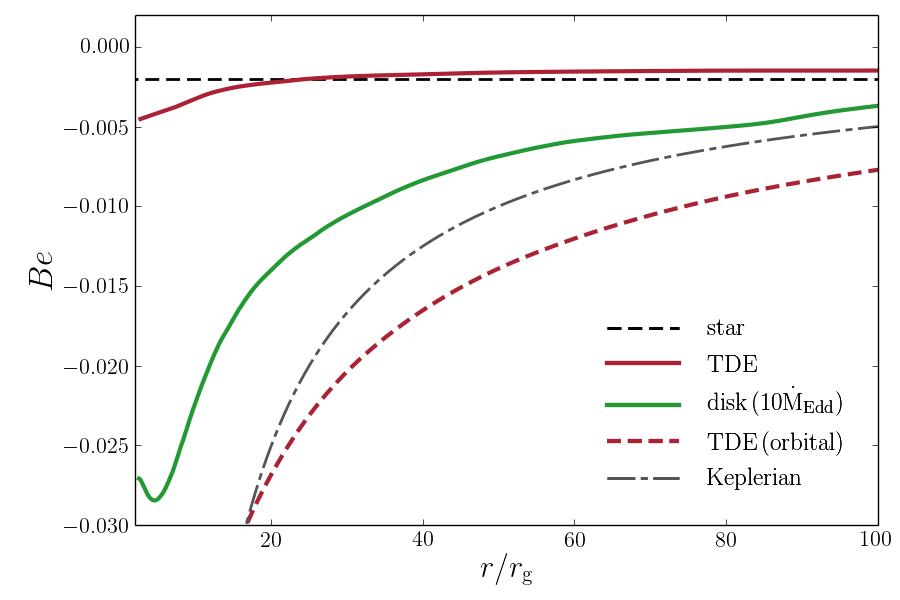}
\caption{Radial profiles of the average Bernoulli function (specific energy) for the
  hydrodynamical simulation of a tidal disruption (solid red line) and a
  typical mildly-supercritical accretion disc (solid green line). The dashed red line reflects
  the orbital energy component of the Bernoulli function of gas in the
TDE simulation. For comparison the specific energy of the star and the
Keplerian orbital energy profiles are shown with black dashed and
dot-dashed lines, respectively.}
\label{f.Bevsr}
\end{figure}

\begin{figure}
\hspace{.3cm}\includegraphics[width=.92\columnwidth]{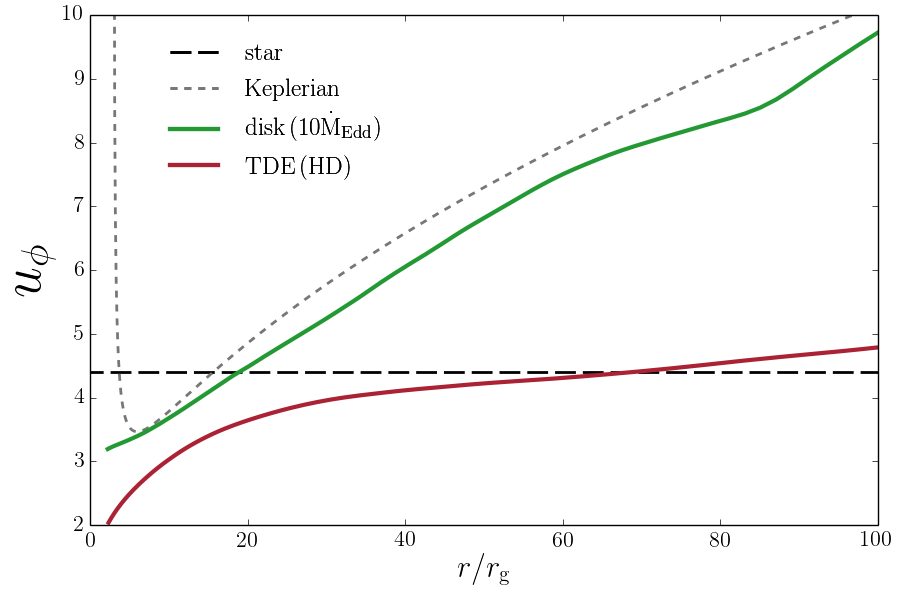}
\caption{Radial profiles of the average specific angular momentum
  in the TDE simulation (red line) and a typical geometrically and
  optically thick accretion disc (green line). The angular momentum of
the stellar orbit and the Keplerian profile are also shown.}
\label{f.Lvsr}
\end{figure}

Figures~\ref{f.Bevsr} and \ref{f.Lvsr} show the average radial
profiles of the specific energy (or Bernoulli function,
Eq.~\ref{e.Be}) and specific angular momentum. The specific energy of
the gas in the circularized disc stays very close to the binding
energy of the disrupted star. This means that the remaining gas 
did not interact strongly enough in the early phases, in
contrast to the gas that was involved in the self-crossing shock
heating, which either got energetic enough to fly out of the domain,
or become bound so much that quickly ended under the BH horizon. 

Having at the same time almost marginally bound gas and a circularized disc seems
contradictory. Marginally bound gas should follow very eccentric
trajectories, not circular orbits, but the situation is complicated by the large thermal
pressure that dominates the energy budget. Due to the low angular momentum,
the orbital energy of the gas in the disc (dashed red line in Fig.~\ref{f.Bevsr}) is 
even lower than the Keplerian value, though the additional contribution of the 
the thermal component ($p/\rho$) allows the debris to be only weakly bound (solid red line).

The specific angular momentum profile for the circularized debris is shown in
Fig.~\ref{f.Lvsr}. In contrast to a typical disc, the specific angular
momentum in the debris is almost constant outside of $r\approx 30\,\rg$ and
very close in value to the angular momentum of the original stellar orbit. This, again, is
the region filled with gas that weakly interacted in the
early stage of the simulation. At all radii, the angular momentum is
sub-Keplerian. Therefore, maintaining equilibrium requires an extra force
compensating the missing centrifugal force. In our case, it is the
gradient of pressure that supports the disc.

To sum up, the gas relatively quickly circularizes and maintains
quasi-stationary, turbulent equilibrium. The disc consists of
weakly bound gas with very low angular momentum. High specific energies
imply a very large geometrical thickness. In many aspects, the
circularized debris resembles the zero-Bernoulli accretion model of
\cite{zebra+14} and agrees with predictions of earlier models of hot
accretion flows \citep[e.g.,][]{narayanyi-94,abramowicz+adafs,adios}.

\subsection{Direct accretion and fallback}
\label{s.fallback}

In this Section we look closely into the rate at which gas crosses the
BH horizon, i.e., at the BH accretion rate. One can distinguish two
phases in the evolution of the tidal debris. Initially, as discussed above, 
the evolution is strongly affected by the shock-heating that develops
at the surface where the deflected debris hits the stream of returning
debris. This process leads to the exchange of energy and momentum
between the streams. The returning stream heats up and drives a
quasi-spherical outflow. Parts of the deflected stream, however,
become more bound and lose angular momentum. As a result, this gas gets
closer to the BH and is likely to cross the horizon. This
\textit{direct accretion} is the main process driving the BH accretion
in the early phase of evolution, while the self-interaction takes
place.

As the disc circularizes, rotating gas reaches a state close to
equilibrium with the gravitational force balanced by pressure gradient
and centrifugal forces. At this stage, direct accretion is no longer possible. 
To be accreted on to the BH, the gas must now get rid of its angular
momentum, though there is no longer a shock surface that could facilitate
that. In standard accretion flows, it is viscosity that
transports angular momentum outwards. This effect indeed occurs in our
simulation, as will be discussed in more detail in Section~\ref{s.viscosity}, but the
accretion driven by viscosity is not the dominant one.

\begin{figure}
\centering\includegraphics[width=.9\columnwidth]{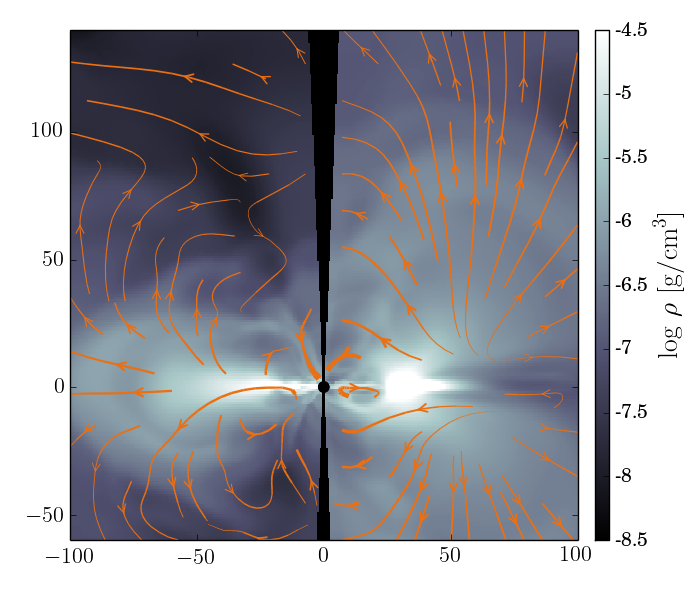}\vspace{-.65cm}
\centering\includegraphics[width=.9\columnwidth]{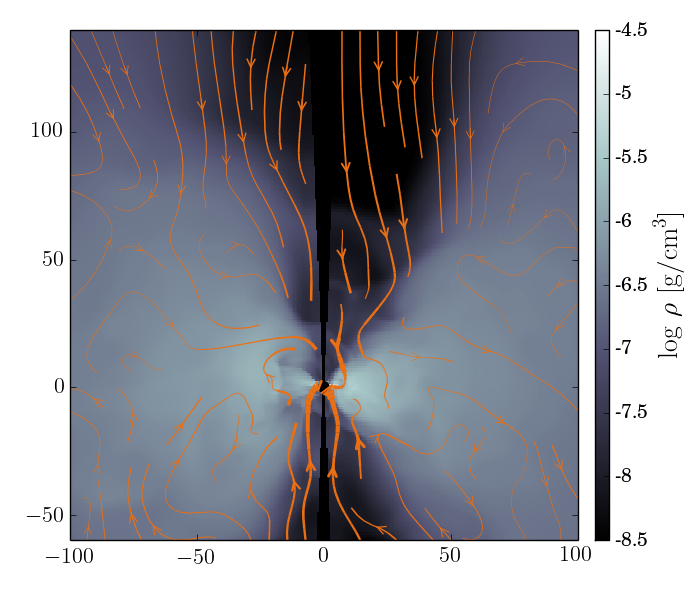}\\
\centering {\vspace{-.3cm}$x/\rg$}
\caption{Distribution of density at the slice through the poloidal
  plane along $y=0$ at $t=10000\,\tg$ (top) and $t=80000\,\tg$ (bottom
  panel). The streamlines reflect the velocity field at the same plane.}
\label{f.fallback}
\end{figure}

Much stronger accretion takes place close to the axis, along the edges
of the rotating disc (see bottom panel of
Fig.~\ref{f.rhomdot}). Fig.~\ref{f.fallback} compares the gas dynamics
in the poloidal plane at the early, self-crossing stage
($t=10000\,\tg$, top panel) with the properties at the late,
after-circularization stage ($t=80000\,\tg$, bottom panel). As long as
the self-crossing shock is present, a significant amount of gas is
driven out of the equatorial plane, and some fraction of it actually
manages to reach the boundaries of and leave the computational domain. The
remaining gas is slowed down by gravity and gradually falls down
towards the hole. The portions of gas that were expelled with low
angular momentum have a chance to come back almost directly to the BH,
ultimately crossing the event horizon, as can be seen from the
bottom panel.

\begin{figure}
\includegraphics[width=.95\columnwidth]{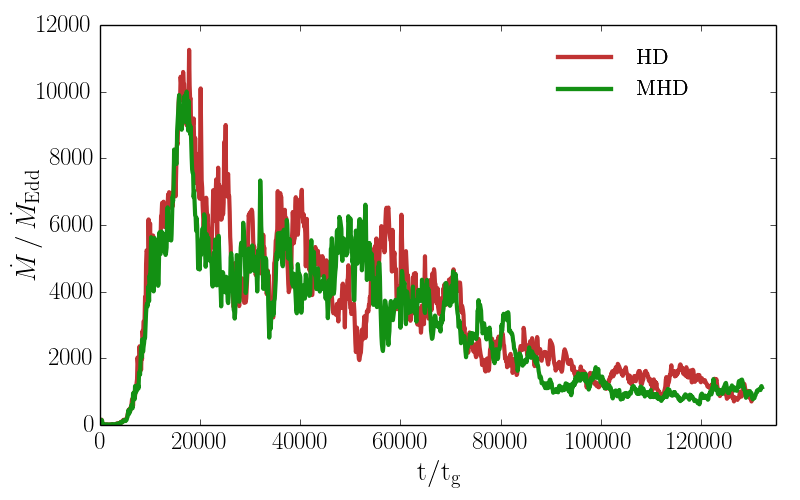}\vspace{-.5cm}
\includegraphics[width=.95\columnwidth]{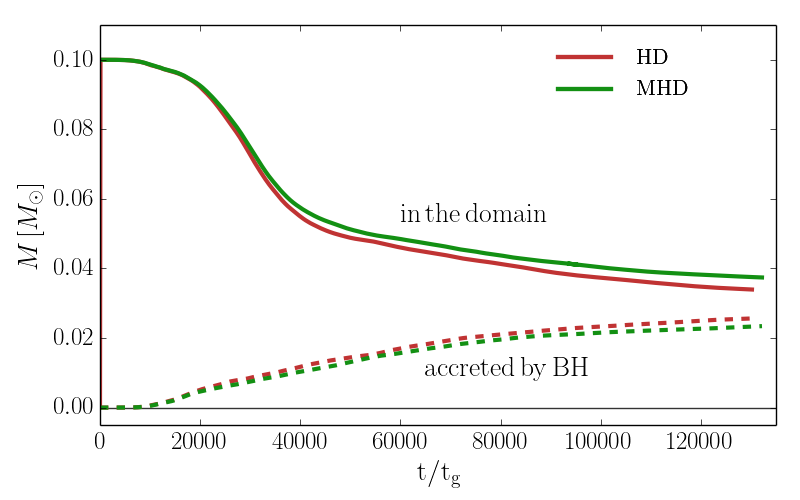}
\caption{Top panel: History of the accretion rate through the BH
  horizon for the hydrodynamical (red) and magnetohydrodynamical
  (green line) simulations. Bottom panel: The amount of mass in the
  computational domain (solid lines) and the mass that has been
  accreted by given time (dashed lines) for the same simulations.}
\label{f.mdots}
\end{figure}

The transition between the two mechanisms driving the accretion is
smooth. The top panel of Fig.~\ref{f.mdots} shows the accretion rate
history through the BH horizon, calculated as
\be 
\label{e.mdot}
\dot M = \int_{0}^\pi
\int_0^{2\pi}\sqrt{-g}\,\rho u^r\,{\rm d}\varphi {\rm d}\theta, 
\ee 
where $\rho$ and $u^r$ stand for density and radial component of four-velocity.
 The peak values are obtained during
the self-crossing phase and can reach as much as $10000\,\Medd$. This
phase dominates until $t\approx 20000\,\tg$. At later stages, the
fallback of the debris ejected out of the equatorial plane becomes
more important, but the transition is not evident. The rate of
  fallback decreases with time and one may expect that at some point
  the viscosity driven accretion will become dominant. The net accretion rate
through the horizon decreases with time and at the end of the
simulation reaches ca.~$1000\,\Medd$ and is still dominated by the
fallback, which is, however, stronger than the viscous accretion only by a
factor of $2-3$ (Section~\ref{s.viscosity}).

The bottom panel of Fig.~\ref{f.mdots} shows the mass contained in the
domain (solid lines) and the mass accreted by the BH (dashed
lines). In the early stage of the simulation ($10000<t/\tg<20000$) the
amount of gas in the domain drops faster than the amount of gas
accreted by the BH increases. The difference reflects the rate at
which the outflowing gas leaves the computational domain. Once the gas
circularizes ($t>20000\,\tg$), gas ceases to flow out through the outer edge and
all the gas lost from the domain ends up inside the BH. By the end of the
simulation, $20\%$ of the star (i.e., $0.02\msun$) has been accreted
by the BH. At the same time, roughly $40\%$ of the stellar gas content
stays in the domain, mostly in the circularized disc. The difference,
i.e., another $40\%$, is the amount of gas that has left the
computational domain. 

\section{Magnetized star case}
\label{s.mhd}

\subsection{Initial setup}
\label{s.mhd.initial}

In addition to the purely hydrodynamical simulation described above,
we performed another one, in which we also evolved the magnetic field. 
The two simulations used exactly the same initial setup, but for
the extra non-zero initial magnetic field inside the stellar debris in
the magnetohydrodynamical one.

The configuration of the magnetic field in the interior of a star is
not well established. Moreover, a tidal disruption introduces additional
complexity. If the stellar magnetic field were known, then one should
follow its evolution throughout the whole disruption, until the gas
falls on to the BH. Recently, an initial effort in this direction has been
made by \cite{guillochon+magn}. However, the SPH code used in our work is
not capable of evolving magnetic fields. Neither do we pretend to know
how to describe the magnetic field inside a star. Instead, we choose
to put a somewhat arbitrary and weak magnetic field inside the stellar
debris only at the time when we translate the SPH output on to the
\koral grid. As long as this magnetic field is weak one can expect
that the field that develops in the ultimately circularized debris
will not depend on the initial conditions, but will rather be
determined by the dynamics of the disc.

We set up the magnetic field at the onset of the grid-based simulation
by setting the (only non-zero) azimuthal component of the vector potential to be
proportional to gas density,
$A^\phi\propto \rho$. The magnetic field is then obtained by
calculating the curl of $A^i$ and is thus guaranteed to be divergence-free. 
The magnitude of the field is then rescaled so that the maximal
magnetic to gas pressure ratio, $\beta'$, does not exceed
$\beta'=0.05$. In this way we obtain a magnetic field that is restricted to the
poloidal plane, forming single loops preferentially in the densest regions
of the stellar debris.

\subsection{Growth of magnetic field}
\label{s.Bgrowth}

When the stellar debris approaches the BH it elongates (although, in
the case of an elliptical orbit like ours, this effect is limited), compresses
and shocks. As a result, the temperature of the debris in the
innermost region grows. The magnetic field does not undergo a similar
increase. Thus, when the debris that remained in the domain
circularizes, the ratio of the magnetic to gas pressures is very low,
$\beta'\approx 10^{-7}$, much lower than in the stellar debris at the
beginning of the grid-based simulation.

The circularized disc shows significant differential rotation. One can
expect that in the presence of non-zero gradient of angular velocity,
radial magnetic field lines will be twisted and stretched along the 
azimuthal direction. The kinetic energy of the rotating flow will be
converted into magnetic energy.

The magnetic field in a differentially rotating flow is unstable against the magnetorotational
instability \citep[MRI;][]{balbus+mri,wielgus+mri}. When the magnetic field strength
is large enough, the induced turbulence starts affecting and dominating the
global dynamics and leads to the dynamo effect. The saturated state is determined by the rate of
shear and non-linear properties of MRI. 

The magnetic field must reach some critical strength
before the MRI can be resolved in a numerical simulation.  The suitable criterion requires that the
fastest growing wavelength (which depends on the strength of the
magnetic field) must be covered by a reasonable ($\sim10$) number of
cells. Therefore, in practice, the magnetic field initially grows only
due to the
differential rotation, and only at some point the MRI starts to be resolved.

\begin{figure*}
\includegraphics[width=.2505\textwidth]{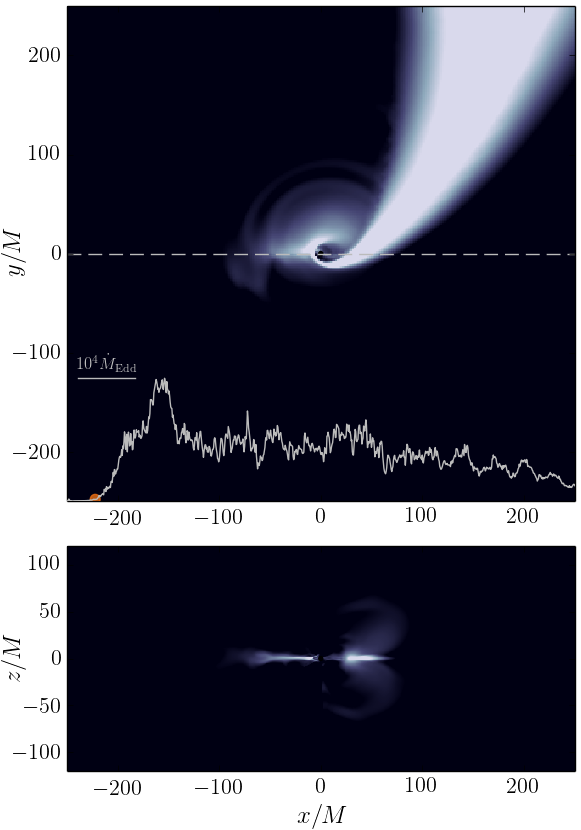}\hspace{-.0cm}
\includegraphics[width=.2245\textwidth]{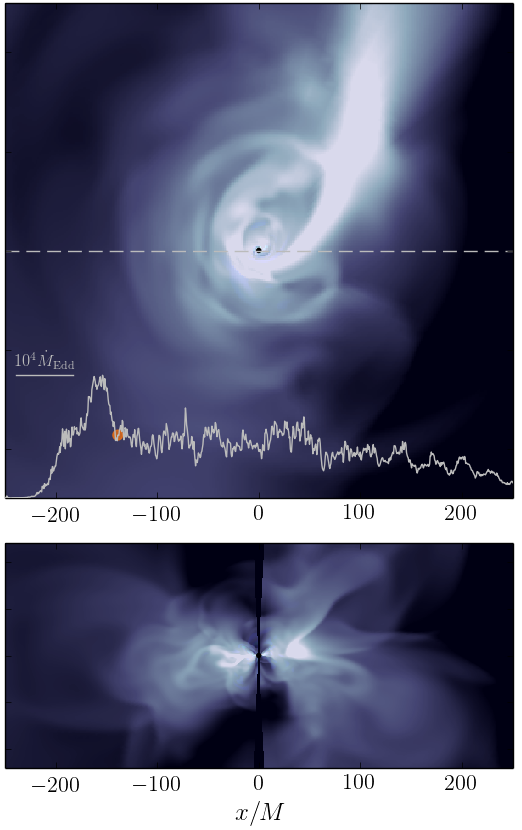}\hspace{-.0cm}
\includegraphics[width=.2245\textwidth]{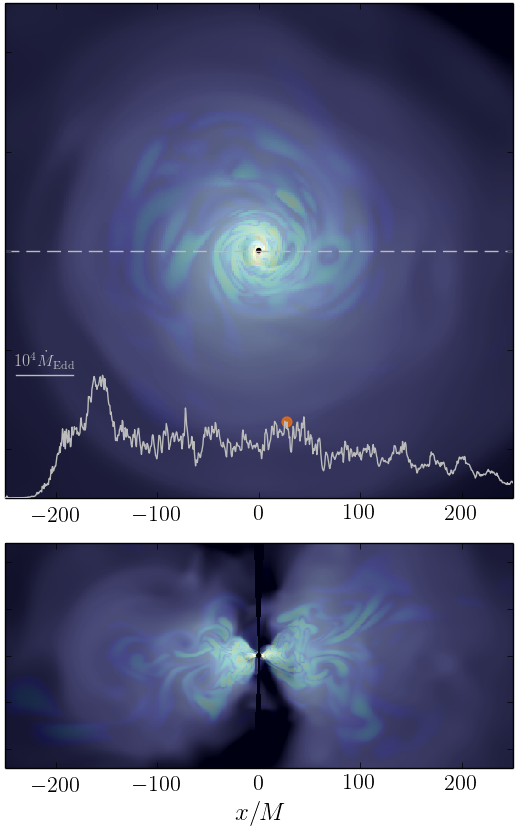}\hspace{-.0cm}
\includegraphics[width=.27\textwidth]{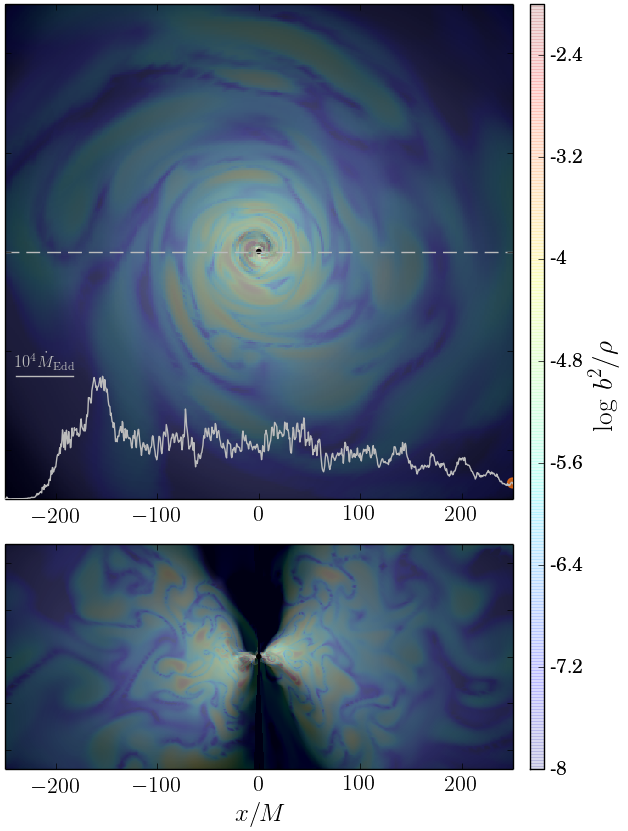}\hspace{-.0cm}
\caption{Similar to Fig.~\ref{f.hd.tde} but for a magnetized
  star. Grey colors reflect the gas density while the colors show the
  magnitude of the, increasing with time, magnetic field. }
\label{f.mhd.tde}
\end{figure*}

Figure~\ref{f.mhd.tde} shows four snapshots of the
magnetohydrodynamical simulation shown in a similar fashion to
Fig.~\ref{f.hd.tde}. The top and bottom panels reflect the equatorial
and poloidal slices, respectively. The density profiles are shown with
grey colors, as before. This time, however, the magnetic field
strength (here measured as ratio of magnetic to rest-mass energy
densities) is plotted on top of the density profiles with colors. The
magnetic field is unimportant for the dynamics in the early phases of
the evolution, i.e., during the self-interaction of the debris. Only
once the disc circularizes, the magnetic field grows in strength on a
dynamical timescale. It builds up the fastest in the innermost region,
where the orbital time is the shortest.

\begin{figure}
\includegraphics[width=.95\columnwidth]{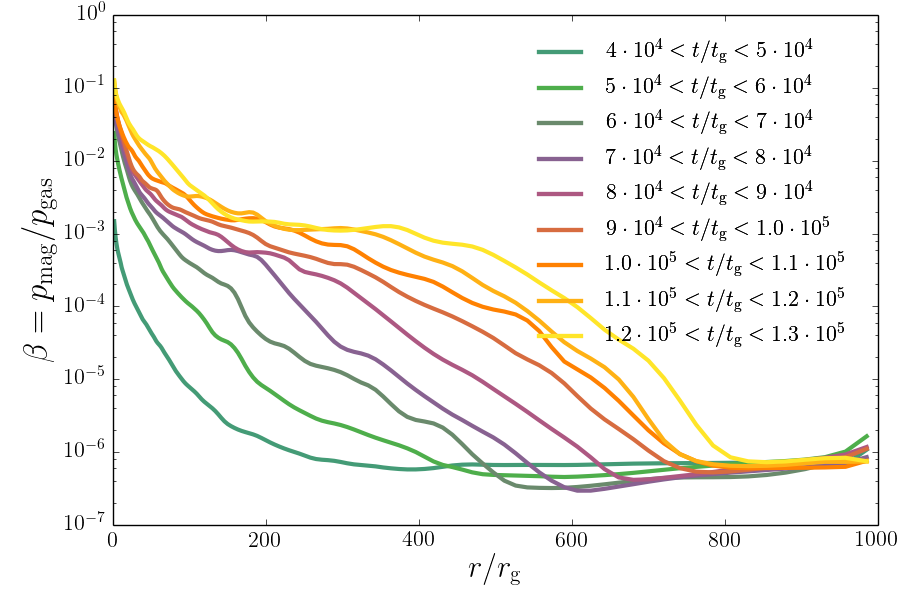}
\caption{Radial profiles of the magnetic to gas pressure ratio,
  $\beta$ in the circularized debris. Lines correspond to the chunks of
time specified in the legend with the green and yellow being the
earliest and latest, respectively.}
\label{f.mhd.beta}
\end{figure}

Figure~\ref{f.mhd.beta} shows radial profiles of the pressure ratio
(magnetic to gas) in
the circularized debris 
obtained by averaging over azimuth and chunks of time lasting $\Delta
t=10000\,\tg$.
The magnetic field initially grows because of the differential
rotation that winds up the magnetic field.  The timescale for
this growth increases with the orbital period, and the magnetic field
grows fastest in the inner region.
Once enough magnetic power has built up to resolve it, the 
MRI grows exponentially and the magnetic field saturates at the level
determined by the non-linear properties of MRI.

\begin{figure}
\includegraphics[width=.95\columnwidth]{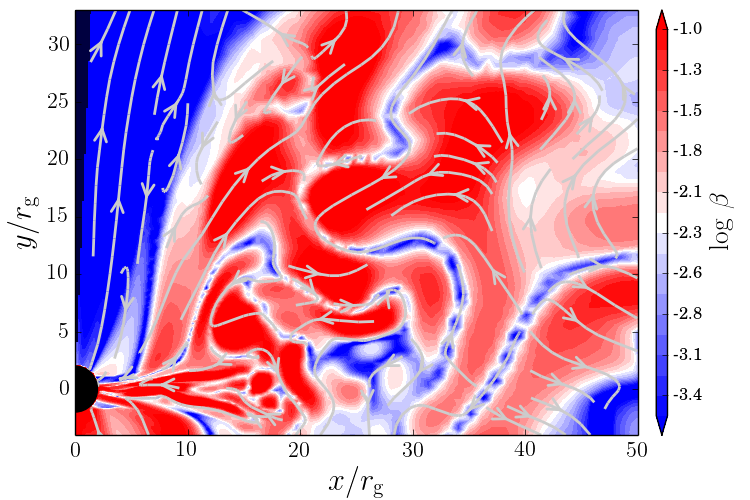}\vspace{-.4cm}
\includegraphics[width=.95\columnwidth]{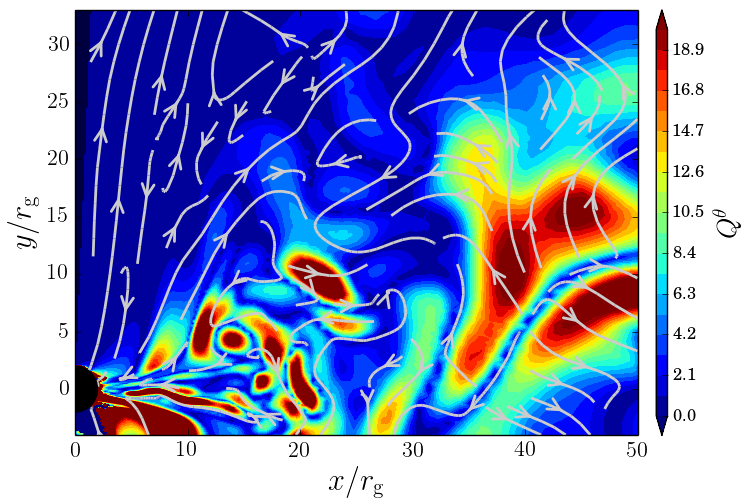}
\caption{Top panel: Magnetic to gas pressure ratio at a slice through
  poloidal place at $t=120000\,\tg$. Bottom panel: The MRI resolution
  parameter $Q^\theta$ (Eq.~\ref{e.qtheta}) plotted in the same
  way. The arrows reflect the poloidal component of the magnetic field. }
\label{f.mhd.betaqtheta}
\end{figure}

Figure~\ref{f.mhd.betaqtheta} shows the magnetic field properties near
the end of the simulation, at $t=120000\,\tg$. The top panel shows the
magnetic to gas pressure ratio at a slice through the poloidal
plane. The magnetic pressure contributes significantly to the total
pressure. In most of the disc interior its contribution exceeds $10\%$
of the total pressure, a value typical for accretion discs
\citep[e.g.][]{hawley+95,penna+alpha}. The bottom panel shows the MRI resolution
parameter $Q^\theta$ defined as \citep{hawley+11,hawley+13},
\be
\label{e.qtheta}
Q^{\theta}=\frac{2\pi}{\Omega \Delta x^\theta}\frac{|B^\theta|}{\sqrt{\rho}}\sqrt{g_{\theta\theta}},
\ee
where $\Delta x^\theta$ is the grid cell size in $\theta$, and $\Omega$ is
the angular velocity. Wherever the magnetic field is oriented mostly
in the polar direction, the MRI resolution parameter reaches its largest
values, significantly exceeding $Q^{\theta}=10$. The MRI seems to be
reasonably well resolved and one may expect that the magnetically
driven turbulence will be present in the simulated flow. However, in
contrast to standard accretion flows, it is not dominant here.

\subsection{Viscosity}
\label{s.viscosity}

The initial phase of the disruption, i.e., the self-interaction of the
streams, is not affected by the magnetic field -- which is still relatively
weak at that stage. Only once the gas forms the quasi-disc,
differential rotation amplifies the field, and magnetic field
saturates at the level specific to MRI, may one expect the
magnetohydrodynamical simulation to differ form the hydrodynamical
counterpart. However, this is not the case. As comparison of
Figs.~\ref{f.hd.tde} and \ref{f.mhd.tde} suggests, the dynamical
properties of the gas do not change. Similarly, the accretion rate
through the BH horizon follows the same time dependence
(Fig.~\ref{f.mdots}). In this Section we study how important the
magnetic field is in sustaining the turbulence and driving the accretion.

The efficiency of a turbulent medium in transporting the angular
momentum, and at the same time in driving the gas inwards, can be
estimated by calculating the effective turbulent $\alpha$ viscosity
parameter. For this purpose we calculate,
\be
\alpha = \frac{\avg{\widehat T^{r\phi}}}{\avg{p}},
\label{e.alpha}
\ee
where $\avg{\widehat T^{\hat r\hat \phi}}$ is the average orthonormal $r,\phi$
component of the stress-energy tensor in the fluid frame, and
$\avg{p}$ is the average gas pressure.

\begin{figure}
\includegraphics[width=.95\columnwidth]{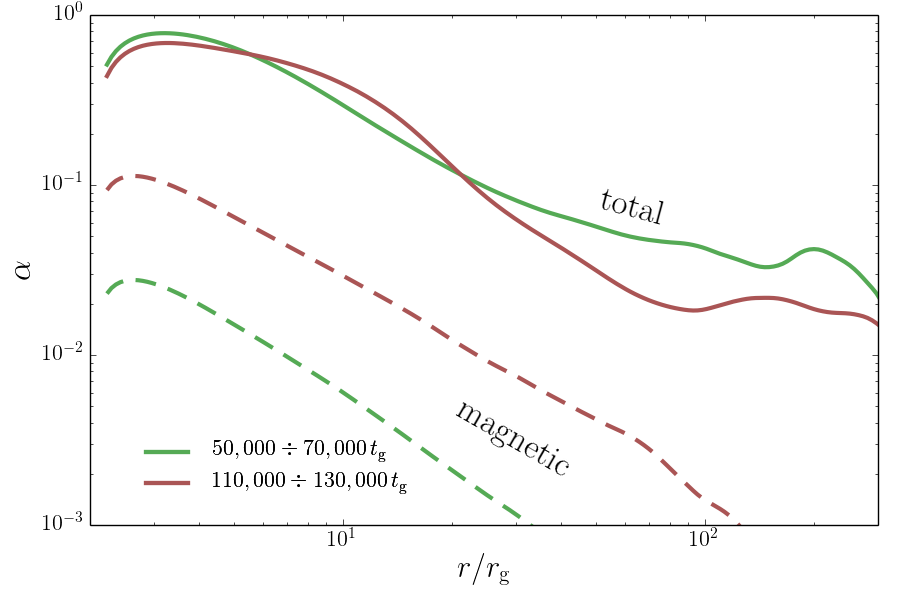}\vspace{-.4cm}
\includegraphics[width=.95\columnwidth]{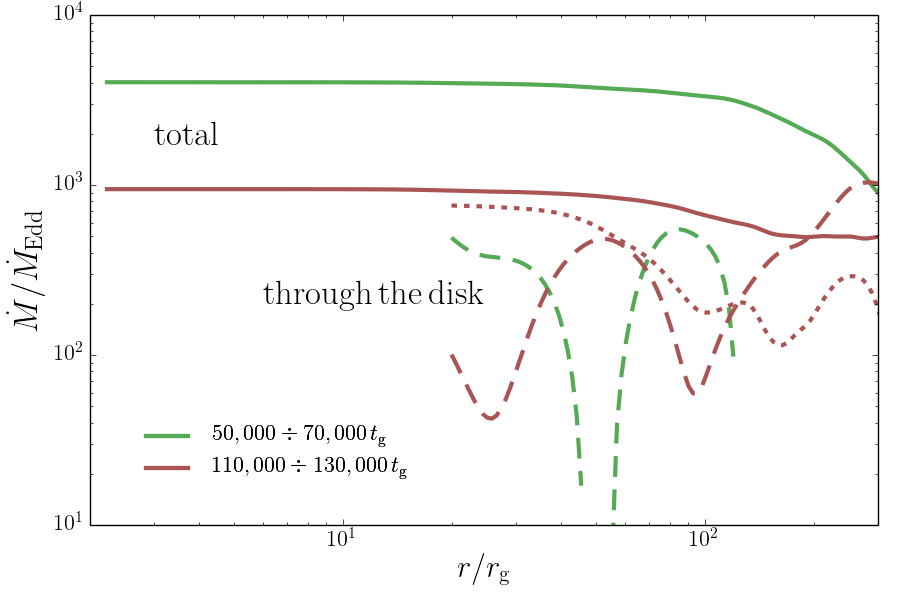}
\caption{\textit{Top panel:} The effective viscosity parameter $\alpha$
  (Eq.~\ref{e.alpha}, solid lines) estimated at two stages of the
  simulation. The dashed lines show the Maxwell component of the
  former. \textit{Bottom panel:} Radial profiles of the mass accretion
  rate integrated over the total domain (solid lines) or only inside
  $\pi/4$ wedge near the equatorial plane (dashed lines). The dotted
  line shows the estimate of the viscous mass transfer rate based on
  the $\alpha$ formalism (Eq.~\ref{e.uralpha}).}
\label{f.mhd.alphas}
\end{figure}

The profiles of the viscosity
parameter at different times calculated within $\pi/8$ from the
equatorial plane for the magnetohydrodynamical simulation
 are plotted in the top panel of  Fig.~\ref{f.mhd.alphas}. The
 solid lines reflect the total stress-energy component, i.e., coming
 from both the hydro- and magnetohydrodynamical components. The
 effective $\alpha$ parameter is quite significant and does not evolve
 significantly with time. It reaches
 $\alpha\approx 0.4$ at $r=10\,\rg$ and increases even further
 towards the BH. Such values of the effective viscosity parameter are
 significantly larger than for regular accretion flows, which exhibit 
 $\alpha\approx 0.1$ at $r=10\,\rg$ \citep{penna+alpha, sadowski+dynamo}.
Does the strong effective viscosity in our simulation come from the
magnetic fields?

The total stress-energy tensor can be decomposed into the Reynolds,
$T^{r\phi}_{\rm Rey}=(\rho+\Gamma u_{\rm int})u^r u^\phi$, and Maxwell
(magnetic), $T^{r\phi}_{\rm Max}=b^2u^r u^\phi-b^r b^\phi$,
component. The comparison between the two allows us to estimate how
important the magnetic field is in transporting the angular
momentum. The dashed lines in the top panel of Fig.~\ref{f.mhd.alphas}
show the Maxwell component of the viscosity parameter. The magnetic
$\alpha$ increased with time, as the magnetic field was building
up. It saturated at the level of $\alpha\approx 0.03$ at $r=10\,\rg$,
which is roughly an order of magnitude less than the  total viscosity
parameter. This fact proves that the magnetic fields do not
contribute significantly to the angular momentum transfer, which is
dominated by the hydrodynamical turbulence left over from the initial
violent self-interaction phase and sustained either by convection or by
gas constantly falling back on the circularized debris. Similar
disproportion of the Maxwell and Reynolds stresses was found in
some of the thick accretion flows simulated by \cite{jon+bigpaper}.

\subsection{Accretion}

Because the impact of the magnetic field on the effective viscosity in the
quasi-disc is very weak, one may expect that the accretion
of gas into the BH will resemble the properties seen before in the
hydrodynamical simulation. This is indeed the case. Initially, gas
falls on to the BH as a result of self-interaction. This stage is
followed by the fallback of gas that has been previously ejected from
the innermost region. It turns out that this mode of accretion
dominates until the very end of both simulations, and that accretion
takes place at similar levels (Fig.~\ref{f.mdots}).

The rate at which the ejected gas falls back on to the BH decreases with
time, as less dense gas, ejected to larger distances, returns. 
As we have shown in the previous Section, the gas in the disc is
turbulent and the turbulence by itself transfers angular momentum
outwards and makes the gas fall towards the BH. The efficiency of
turbulent hydrodynamical viscosity does not seem to decrease with
time (see top panel of Fig.~\ref{f.mhd.alphas}), so one may expect
that at some point the accretion of gas contained in the disc will
overcome the continuously decreasing fallback rate.

We estimate the accretion rate mediated by the turbulence by
calculating the mass flux inside the bulk of the disc, outside the
innermost region where the flow is strongly affected by the debris
falling back, and comparing it with the total accretion rate. The rates
are calculated at various radii following Eq.~\ref{e.mdot}, but
limiting the range of polar angles for the former to stay within $\pi/8\,\rm rad$
from the equatorial plane, which is enough to capture most of the
accretion taking place there (compare Fig.~\ref{f.rhomdot}).

The bottom panel of Fig.~\ref{f.mhd.alphas} shows the radial
  profiles of the average accretion rate calculated at two stages of the
  simulation by averaging in time and azimuth. The solid lines reflect the total
  mass flux. They are flat up to $r\approx 50\,\rg$, reflecting the
fact that the flow has reached the equilibrium state within this
region. However, the rate at which the gas flows on to the BH decreases 
with time, as a result of the decreasing fallback rate.

Dashed lines in the same plot show the accretion rate inside the
  bulk of the debris. They are not as flat as the profiles for the
  whole domain because of two reasons. First, the radial velocities
  in this region are much lower than the radial velocities of the
  infalling gas, and it takes longer time to reach
  equilibrium. Second, the  gas is not guaranteed to stay inside the
  region of integration (as it would be required in order to obtain a constant
  accretion rate profile). Nevertheless, it is possible to approximate
  the accretion rate driven by turbulent viscosity in the
  bulk of the debris. It falls between $50$ and $500\,\Medd$, a
  range that does not significantly change with time, in agreement with the
  constancy of the effective viscosity parameter, $\alpha$, discussed
  previously.

The turbulence-driven accretion rate may also be estimated using
  the analytical expression for the radial velocity of an accretion
  flow \citep[e.g.,][]{frankkingraine85},
\be
u^r\approx \alpha (h/r)^2 \Omega r,
\label{e.uralpha}
\ee
 where $h/r$ is the
scale height of the disc and $\Omega$ is the angular velocity. The
related accretion rate is given through, $\dot M=2\pi r \Sigma u^r$,
where $\Sigma$ denotes the surface density. Reading radial profiles of
$\alpha$ and $\Sigma$ from the simulation, estimating $h/r\approx 1$
and calculating the radial velocity using Eq.~\ref{e.uralpha} we
obtain the accretion rate profile plotted with dotted line in the
bottom panel of Fig.~\ref{f.mhd.alphas}. It has the same order of
magnitude as the rate estimated by direct integration of the mass flux
inside the debris disc, proving that it is indeed the turbulence
that drives the gas towards the BH. However the rate is lower than
the fallback rate even at the end of the simulation, and one would have
to wait even longer for the gas driven in this way to dominate the mass flux
through the horizon.

\section{Discussion}
\label{s.discussion}

\subsection{Optical depth and luminosity}

The simulations described in this paper were performed within the
framework of general relativistic hydro- and magnetohydrodynamics. The
close elliptical disruption that we simulated led to a moderately
elongated tidal debris stream returning to the BH after the first
pericenter passage. As a result, the optical depths of both the debris and 
the subsequent circularized disc were very large. The optical
depth due to Thompson scattering across the returning debris
(Fig.~\ref{f.sph}) was of the order of $10^7$. The circularized debris
has the optical depth only an order of magnitude smaller, $\tau\approx
{\rm few} \times 10^5$. With such a large amount of  gas spread
throughout the computational
domain, the photosphere is not located within its boundaries. In the initial
phase of violent self-interaction, the outflow itself is optically
thick. In the late, after-circularization stage, the density near the
edge of the box is roughly $10^{-10}\,\rm g/cm^3$, high enough to
provide an optical depth of unity over one gravitational radius,
$r_{\rm g}\approx 1.5\cdot 10^{10}\,\rm cm$.

With the photosphere effectively outside of the computational domain,
it made no sense to run an analogous simulation with radiative
transfer incorporated (although the \koral code is capable of
that). As long as the gas is very optically thick and the photons do
not have an opportunity to detach from the gas, its evolution may be
followed with pure hydrodynamics, with an appropriate adiabatic index 
$\Gamma$. 

However, not resolving the photosphere and evolving the debris
as optically thick within the hydrodynamical framework does not allow
for direct measurement of the radiative luminosity emerging from the
system. To do it properly, one would have to evolve the radiation
field in parallel and make sure that the photosphere, where photons
decouple from gas, is within the domain. Therefore, we cannot measure
the light curve directly.

However, extracting the radiative efficiency would be of interest. The
observed tidal disruption events seem to emit less radiation than
expected from thin disc accretion of the amount of stellar debris that
is bound to the BH, and the reason for this is still unclear
\citep[see, however,][]{piran+15b}. We can estimate the luminosity
extracted from the system by calculating the flux of thermal energy
crossing the outer edge of the domain. In radiation pressure
dominated gas, this quantity would approximately correspond  to
the amount of radiative energy carried by photons advected with the 
optically thick gas. If no further interaction between gas and
radiation takes place outside the boundary, it would also reflect the
flux of energy carried away by free-streaming photons or, after
integrating over the whole sphere, the total radiative luminosity of
the system.

\begin{figure}
\includegraphics[width=.95\columnwidth]{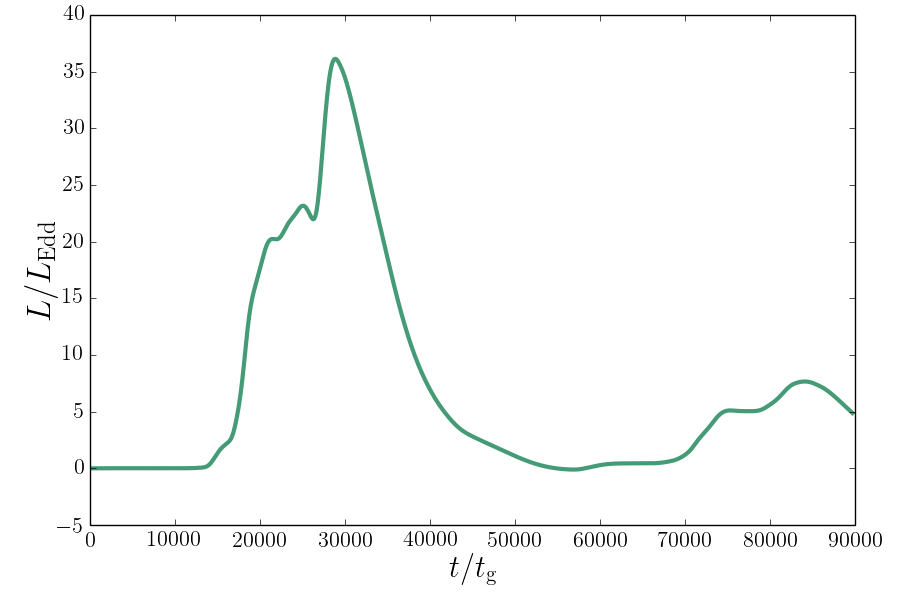}
\caption{Luminosity (flux of energy integrated over the whole sphere) as a
  function of time. The luminosity was estimated by taking the
  advective flux of internal energy crossing the outer edge of
  computational domain (see Eq.~\ref{e.Lum}). }
\label{f.lumvst}
\end{figure}

Our estimate of the radiative luminosity is calculated as,
\be
L=\int_0^{2\pi}\int_0^\pi u_{\rm int}\,u^r\sqrt{-g}\,{\rm d}\theta {\rm
  d}{\phi},
\label{e.Lum}
\ee
where $u_{\rm int}$ is the internal energy of the
gas. Fig.~\ref{f.lumvst} shows the corresponding light curve for the
hydrodynamical simulation. The
luminosity increases and reaches maximum when gas flows out through
the edge of computational domain, i.e., between $20000$ and
$40000\,\tg$. This period of time corresponds to the largest flux of
thermal energy crossing the boundary and reflects the largest flux of
radiative energy carried with optically thick, radiation pressure
supported gas. However, even during this period, the luminosity is only
ca.~$30\,\ledd$, much less than the outflow accretion rate ($\sim
40000\,\medd$). Once the outflow ends (which happens when
the self-crossing shock is no longer present), the luminosity goes
down and hardly any radiation is carried out by the gas. However, our
estimate does not include the diffusive flux of radiation, which in
principle can be present even in stationary gas. This component,
however, will never exceed $\ledd$ in optically thick gas.

 Summing up,
one can expect a short episode of increased, but moderate, luminosity
when gas is ejected during the self-crossing phase. It is followed by
a low luminosity despite the fact that the BH is actually accreting
at a highly super-Eddington level. The accretion is very inefficient,
mostly because gas falling on the BH is almost marginally bound
  and does not extract significant energy,
preventing the production of significantly super-Eddington luminosities.

\subsection{The self-crossing feedback loop}

In Section~\ref{s.periodicity} we have discussed the feedback loop
that takes place when the debris returns on a close orbit towards the
BH, is deflected due to the apsidal precession, and hits the original
stream at an angle. The momentum transferred in this way pushes the
returning stream outwards, decreasing the efficiency of the feedback
exerted by the deflected stream, and allowing the incoming stream to
revert to its original trajectory. This feedback modulates the
self-crossing and the related dissipation in shocks. It may also
results, like in our case,  in modulating the rate
at which
outflows are driven out of the self-intersection region. 

This
modulation is visible in Fig.~\ref{f.inout}, where the grey line
reflects the rate at which outflowing gas crosses the sphere of
radius $r=100\,\rg$. The red line in the same plot shows the
corresponding rate as measured at $r=1000\,\rg$ (i.e., at the edge of the
domain). The periodical imprint is no longer visible, since
the modulation of the outflow generation rate results in ejection of consecutive
shells moving with different velocities. Such shells 
collide with each other inside the domain and the initially
modulated outflow rate averages out to provide a smooth profile at
large radii.

Therefore, although the feedback loop is evident and takes place in
the self-intersection region, one should not expect to see the related
modulation imprinted on the kinetic energy of the outflow leaving the
system, nor on the radiative light curves emerging when
the initially optically thick outflow ultimately becomes optically thin.

However, one has to keep in mind that the tidal disruption simulated
and analyzed in this paper is peculiar because of its ellipticity (although possible sources of 
highly eccentric TDEs have been discussed by \citealp{hayasaki13}), 
which prevents the orbit from becoming elongated, leading to the
returning debris being extremely optically thick. In the case of a more
realistic, parabolic encounter, the debris would consist of gas of
lower density and the resulting optical depth would be lower. In such
a case, the radiation generated and modulated in the shocked region
will decouple from the gas much earlier, and is likely to carry the
imprint of the periodicity out to infinity. The characteristic timescale
will roughly correspond to the Keplerian orbital period at the
self-intersection radius.

\subsection{Extrapolation to parabolic encounters}

The star that underwent the  tidal disruption described in this work
was originally on an eccentric orbit with eccentricity $e=0.97$. Such
a setup was favorable for our simulations -- the
tidal debris was confined to a relatively small volume and could be
translated on to the \koral grid. Similarly, the timescale of the
disruption and the following debris fallback episode were short, making
the computations feasible.

However, such encounters are not very likely. Most stars are directed
towards the BH by dynamical interactions with other objects when they
are very far from the BH. Thus, they approach the central mass
effectively along parabolic orbits.

Parabolic disruptions result in much more elongated streams of
returning debris, with half of the stellar mass actually becoming unbound. 
Therefore, the initial phase of the simulations
discussed here, i.e., the self-crossing of the streams, will last
longer. In the case discussed in this work, the self-interaction ended
when all the debris returned to the BH. In the case of a parabolic
disruption, the mass return rate never formally reaches zero, 
although the amount of mass that falls back will keep decreasing with time. 
One can expect that the self-intersection, as long as it
takes place close to the BH -- where plenty of energy is available for
dissipation--, will dominate the dynamics in the initial phase. However,
when the amount of gas that circularized in the inner region is much
larger than the gas supply rate at some characteristic timescale, the
incoming stream will no longer be dynamically important to the
circularized debris. Thus, as in our case, the self-crossing phase
will be followed by circularization.

All the characteristic features of the close, elliptic disruption
described in this work will hold for a close, parabolic encounter,
although likely with different efficiencies. The self-intersection
will result in a feedback loop, which may modulate the emerging light
curves. Gas will be ejected from the shocked region and will keep
falling back on to the BH for some time. The circularized debris will
form a marginally bound quasi-disc, which is likely to provide only 
moderate or weak turbulent viscosity.

\subsection{Extrapolation to lower impact parameters}

Many of the properties discussed so far are strictly related to the
efficiency of the interaction in the self-crossing shock that forms
due to the relativistic precession. The large impact parameter in our
simulation translates into a small pericentre radius (in terms of $r_{\rm g}$), 
resulting in significant apsidal precession and in the self-interaction taking place close to the BH. 
This fact implies that the colliding streams carry significant kinetic
energy, which can be dissipated and redistributed between the
streams. As a result, a significant fraction of the returning debris
gains energy, and an outflow -- to some extend unbound -- forms. At the
same time, a significant amount of debris falls almost directly on the
BH. The large amount of energy involved in the violent interaction 
makes the flow very turbulent, and this turbulence does not decay for
many orbital periods, providing -- at least initially -- the effective
viscosity needed for accretion  from the debris disc.

One has to be aware, however, that such close encounters as studied in
this work are not common. It is unlikely for a star to approach a
SMBH along an orbit plunging so deeply into the loss cone. What
changes in the described picture if the impact parameter is lower,
i.e., if the star does not come as close to the BH? Most importantly,
the kinetic energy available for dissipation and redistribution in the self-crossing
shock would be lower, leading to less efficient outflow
and less pronounced turbulence. Furthermore, the feedback loop
described in Section~\ref{s.periodicity} would be less efficient, and
the related periodicity would probably disappear. The circularization would
proceed along the standard picture, i.e., without any significant outflows
preceding it and with little intrinsic turbulence, and the accretion
on to the BH would start only after the magnetic field manages to grow and
make the debris turbulent. Therefore, the novel features discussed in
this work, i.e., the outflow, periodicity and hydrodynamical
turbulence, should be regarded as characteristic to close tidal disruptions.

\subsection{Impact of BH spin}

We have performed simulations of a tidal disruption of a red dwarf by
a non-rotating SMBH. A non-zero BH spin would affect two major factors
determining the outcome of the debris evolution. Firstly, as long as
the star is in the equatorial plane of the BH (the
plane perpendicular to the BH spin), a non-zero BH spin would only
have a minor contribution to the deflection angle due to the apsidal precession.  For
BHs rotating against the debris motion (negative BH spin), the self-crossing would take
place closer to the BH,  most likely resulting in more efficient shocks, 
leading to larger outflow and direct inflow rates in the initial
phase. BH rotation along the debris orbit (positive spin) would move
the self-intersection further out, decreasing the efficiency of the
shock-related processes, and in particular their associated radiative
emission.

However, there is no reason to expect the stellar orbit to lie in the
equatorial plane of a spinning BH.  If the spin and
misalignment angle are large enough, Lense-Thirring precession
will change the plane of the debris motion. This will likely result in
the deflected stream missing the original one, which will make the
shocking less efficient, or not present at all. In such a case, there
will be little or no outflow -- radiation from the initial,
self-crossing phase, and the circularization will be delayed. The
orbital and spin plane misalignment may therefore explain the
disagreement between the observed and predicted rates of tidal
disruption events \citep{stone+16}.

\subsection{Comparison with other works}

Our work combines SPH and grid-based methods to study a close tidal disruption
and the following accretion on to the BH. We ran two simulations, with and without magnetic
fields. We assumed zero BH spin, an orbital eccentricity of $e=0.97$, and a
mass ratio $M_{\rm BH}/M_{*}=10^6$. In recent years, a number of
groups performed numerical studies of tidal disruptions and the ensuing 
circularization. Our work is unique because it consistently evolves
magnetic fields and tracks the gas down and across the BH
horizon. Below we briefly summarize simulations most similar to ours.

\cite{shiokawa+15} followed a similar approach, and combined a code
allowing the study of a stellar disruption in the tidal
gravitational field of a BH (in their case it was a grid based code,
with the grid comoving with the star, that accounts for self-gravity and
applies multipole expansion of the tidal gravitational field) with a
grid-based GR hydrodynamical code. The authors studied a disruption of
a white dwarf approaching a $500\,\Msun$ along a parabolic orbit. The
inner edge of their simulations was located at $r=30\,\rg$. They
found tidal streams self-interacting and identified multiple shocks lasting throughout the
simulation. Their results are complementary with ours as they
study a different regime -- the self-interaction in their case takes
place at a much larger distance ($r\approx 1000\,\rg$) where less kinetic
energy can be dissipated. The existence of multiple shocks in the
debris can be explained by the fact that they study a parabolic
encounter with a tidal stream of debris constantly flowing into the
domain and affecting the already circularized debris.

\cite{bonnerot15} studied circularization of the debris using a two-phase 
SPH approach where the initial disruption is studied including
self-gravity, while the following circularization stage is followed without taking it
into account. For both stages
they applied the \cite{tejedarosswog-13} pseudo-Newtonian potential that allows 
studying the impact of relativistic precession. The authors
investigated elliptical orbits ($e=0.8$) and found that the debris stream circularizes, 
after a few orbits, at the pericenter radius. They also showed
that the structure of the debris disc that forms is sensitive to its
cooling efficiency.

All the studies mentioned so far assumed negligible
BH spin. As discussed earlier, misaligned orbital and spin planes can
lead to inefficient self-interaction. This effect was studied by
\cite{hayasaki15}, who performed SPH simulations with  post-Newtonian
corrections of multiple
elliptical disruptions. They have shown that indeed the Lense-Thirring
precession can delay the circularization of the debris.

Similar conclusions were obtained earlier by \cite{haas+12}, who
studied a disruption of a white dwarf by an intermediate mass BH using
a grid-based relativistic hydrodynamical code that solves the
Einstein equations consistently. The authors also claimed that the luminosity of
the debris never exceeds the Eddington luminosity because of photon
trapping in  the super-Eddington accretion flows.

\section{Summary}
\label{s.summary}

We have performed two simulations of a close (impact parameter
$\beta=10$), elliptical disruption of a red dwarf star by a
non-rotating supermassive BH. We used an SPH code to simulate the
initial passage and subsequent disruption of the star, and a
grid-based code to simulate the return of the debris towards the BH.
One of the grid simulations was purely hydrodynamical, while the
other included a magnetic field seeded at a low level, at the very onset
of the grid-based calculations.  Our results are summarized
as follows:

\begin{itemize}

\item \textit{Outflow and direct accretion} -- Due to the high impact
  parameter, the simulated debris
  returns very close to the central BH. The apsidal precession of the 
  orbit results in efficient self-crossing of the streams taking place
   roughly between $r=10\,\rg$ and
  $100\,\rg$. 
  The deflected stream hits the original one and a shock
  forms. The kinetic energy of the former is dissipated and the
  gas is heated up. This interaction redistributes energy and makes
  the deflected gas more bound, while the shocked gas acquires energy 
  and becomes less bound. As a result, the hot and energetic gas, often unbound,
  expands and forms a significant quasi-spherical outflow emerging
  from the self-crossing shock region. In the simulated disruption,
  roughly $40\%$ of the debris left the computational domain ($1000\,\rg$) in this
  way. The gas that lost energy and
  became more bound is no longer able to orbit around the BH and 
  crosses the BH horizon.

\item \textit{Self-crossing related periodicity} -- The deflected
  stream hits the incoming one and pushes it slightly 
  outwards, increasing its periapsis distance. This, in turn, decreases the
  apsidal precession angle and moves the self-crossing region further
  from the BH, decreasing the efficiency of interaction, and allowing 
  the incoming stream to revert to its previous orbit,  once again with an 
  increased deflection angle. 
  This phenomenon results in a
  feedback loop with characteristic period roughly corresponding to the
  Keplerian angular momentum at the region of self-intersection
  (Eq.~\ref{eq.nukepl}). The feedback modulates the dissipation
  taking place in the self-crossing shock. In our
  simulations, the related periodicity was imprinted on the rate of
  optically thick
  outflow generated from the shocked region. However, in more
  realistic parabolic disruptions, which are more optically thin, the
  periodicity may be directly visible in the light curves. A necessary
  condition for the modulation to occur is an efficient
  self-crossing, which in turn requires a large impact parameter,
  and either a very small BH spin or an orbital plane perpendicular
  to the BH spin.

\item \textit{Circularized but turbulent marginally bound debris} -- 
The debris forms a rotating thick disc which on
  average is in equilibrium, but at the same time is turbulent. This
  turbulence is inherited from the initial violent stage of
  self-interaction and does not dissipate for a long time. The debris
  is by itself partially convectively unstable and is constantly
  perturbed by the gas
  falling back after being ejected during the shock phase. Both factors
  help maintain the turbulence. The
  magnetorotational instability (MRI) does not change this picture --
  the debris remains dominated by the original hydrodynamical
  turbulence for a long time. It lasts until the end of the
  simulations we performed, but one may expect that ultimately the
  MRI-driven turbulence takes over, unless the convection manages to
  sustain the observed turbulent power.

\item \textit{The average properties of circularized debris} -- The
  self-crossing of the tidal stream results in shocks that 
  significantly heat up the gas. However, the total specific energy of the gas
    in the inner region
    is close to the specific energy of the stellar orbit (for a
    parabolic orbit, the corresponding Bernoulli function, $Be$, would
    be zero). The high thermal pressure provides a pressure gradient that,
    together with the centrifugal force, balances out the gravitational force.
  As a result, the circularized
  debris is close to marginally bound and follows the general picture
  of the zero Bernoulli accretion flow \citep[e.g.][]{zebra+14}. 

\item \textit{Fallback accretion} -- Only a small fraction of the debris that was
  ejected from the self-crossing shock region is energetically
  unbound. The bound debris gradually decelerates and after a
  while starts falling back towards the BH. The original outflow was
  quasi-spherical. Some fraction of the gas was ejected with low
  angular momentum, which allows it to fall back almost directly on to the
  BH. We find that after the initial phase of self-interaction and
  direct accretion, most of the gas crossing the BH horizon actually
  comes from the debris falling back in the polar region, along the
  edges of the disc. Only near the end of the simulations did the fallback
  rate get close to the estimated turbulence-driven accretion taking place in the
  debris disc. We predict that fallback accretion dominates for an extended period
  of time for any close tidal disruption that results in efficient
  self-crossing near the BH.

\item \textit{The effective viscosity} -- Surprisingly, we find that
  magnetic fields do not change significantly the properties of the
  debris disc.  The turbulence inherited from the initial, chaotic and
  violent self-interaction stage does not decay in either the hydro- or
  magnetohydrodynamical simulations for many orbital periods. The effective viscosity provided
  by this turbulence dominates over the viscosity mediated by magnetic
  fields, and will lead to accretion from the disc once the fallback ceases.

\item \textit{Radiative efficiency} -- The simulated close tidal
  disruption is likely to be extremely radiatively inefficient. The
  initial self-crossing phase led to ejection of optically thick
  outflow carrying significant kinetic energy. However, the amount of
  radiation that is likely to reach an
  observer once that outflow becomes optically thin is not large --
  despite the outflow rate at the level of $40000\,\Medd$, the estimated
  luminosity did not exceed $40\,\Ledd$. This picture could change, however, if
  the kinetic energy of the outflow is dissipated. Neither will the bound
  debris falling back on to the BH lead to significant radiative
  emission -- such gas is optically thick, hardly bound, and falls
  back in a laminar way. Once the fallback ceases, the accretion from
  the debris disc will dominate. The radiative efficiency of this process will be
  lower than for a standard accretion disc because the debris holds to
  the energy of the original star (so that little energy can be
  extracted when the gas approaches the horizon). Furthermore, as
  long as the accretion rate and the optical depth of the debris are
  large, the efficient photon trapping does not allow for large luminosities.

\end{itemize}

\section{Acknowledgements}

The authors thank Ramesh Narayan, Julian Krolik, Chris Fragile,
Jean-Pierre Lasota, James Guillochon and Jonatham McKinney for useful comments.  AS
acknowledges support for this work by NASA through Einstein
Postdoctoral Fellowship number PF4-150126 awarded by the Chandra X-ray
Center, which is operated by the Smithsonian Astrophysical Observatory
for NASA under contract NAS8-03060. AS and EG thank the
Harvard-Smithsonian Center for Astrophysics for its hospitality.  The
work of SR has been supported by the Swedish Research Council (VR)
under grant 621-2012-4870.  The authors acknowledge computational
support from NSF via XSEDE resources (grant TG-AST080026N), from NASA
via the High-End Computing (HEC) Program through the NASA Advanced
Supercomputing (NAS) Division at Ames Research Center, and from
  the North-German Supercomputing Alliance (HLRN).
 
\bibliographystyle{mn2e}

{\small
\begin{thebibliography}{}


\bibitem[Abramowicz et al.(1995)]{abramowicz+adafs} Abramowicz, M.~A., 
Chen, X., Kato, S., Lasota, J.-P., \& Regev, O.\ 1995, \apjl, 438, L37 


\bibitem[Arcavi et al.~(2014)]{arcavi14} Arcavi, I., Gal-Yam, A., Sullivan, M., et al. 2014, ApJ, 793, 38

\bibitem[Balbus 
\& Hawley(1991)]{balbus+mri} Balbus, S.~A., \& Hawley, J.~F.\ 1991, \apj, 376, 214

\bibitem[\protect\citeauthoryear{{Balsara}}{{Balsara}}{1995}]{balsara}
{Balsara} D.~S.,  1995, Journal of Computational Physics, 121, 357

\bibitem[\protect\citeauthoryear{{Benz}, {Cameron}, {Press} \& {Bowers}}{{Benz}
  et~al.}{1990}]{benz90}
{Benz} W.,  {Cameron} A.~G.~W.,  {Press} W.~H.,    {Bowers} R.~L.,  1990,
  Astrophysical Journal, 348, 647
  

\bibitem[Blandford 
\& Begelman(1999)]{adios} Blandford, R.~D., \& Begelman, M.~C.\ 1999, \mnras, 303, L1 


 \bibitem[Bloom et al.~(2011)]{bloom11} Bloom, J. et al. 2011, Science, 333, 203
  
\bibitem[\protect\citeauthoryear{{Bonnerot}, {Rossi}, {Lodato} \& {Price}}{{Bonnerot} et al.}{2015}]{bonnerot15}
{Bonnerot}, C., Rossi, E., Lodato, G. \& Price. D.J.  2015, arXiv:1501.04635  

\bibitem[Brown et al.~(2015)]{brown15} Brown, G.~C., Levan, A.~J., Stanway, E.~R., Tanvir, N.~R., Cenko, S.~B., Berger, E., Chornock, R. \& Cucchiaria, A. 2015, \mnras, 452, 4297

\bibitem[Cenko et al.~(2012a)]{cenko12a} Cenko, S.~B. et al. 2012a, \mnras, 420, 2684

\bibitem[Cenko et al.~(2012b)]{cenko12b} Cenko, S.~B. et al. 2012b, \apj, 753, 77

\bibitem[Chornock et al.~(2014)]{chornock14} Chornock, R. et al. 2014, \apj, 780, 44

\bibitem[Coughlin 
\& Begelman(2014)]{zebra+14} Coughlin, E.~R., \& Begelman, M.~C.\ 2014, \apj, 781, 82 

\bibitem[Coughlin et al.~(2015)]{coughlin15} Coughlin, E.~R., \& Nixon, C.\ 2015, \apj, 808, L11 


\bibitem[Frank et al.~(1985)]{frankkingraine85} Frank, J., King, A.~R., 
\& Raine, D.~J.\ 1985, Cambridge and New York, Cambridge University Press, 1985, 283 p.,  


\bibitem[\protect\citeauthoryear{{Gafton}, {Tejeda}, {Guillochon}, {Korobkin}
  \& {Rosswog}}{{Gafton} et~al.}{2015}]{gafton15}
{Gafton} E.,  {Tejeda} E.,  {Guillochon} J.,  {Korobkin} O.,    {Rosswog} S.,
  2015, MNRAS, 449, 771
  
\bibitem[Gezari et al.~(2008)]{gezari08} Gezari, S. et al. 2008, \apj, 676, 944

\bibitem[Gezari et al.~(2009)]{gezari09}Gezari, S. et al. 2009, \apj, 698, 1367

\bibitem[Glatzel(1988)]{glatzel-88} Glatzel, W.\ 1988, \mnras, 
231, 795 

\bibitem[\protect\citeauthoryear{{Gingold} \& {Monaghan}}{{Gingold} \&
  {Monaghan}}{1977}]{GM77}
{Gingold} R.~A.,  {Monaghan} J.~J.,  1977, Monthly Notices of the Royal
  Astronomical Society, 181, 375
  
\bibitem[Guillochon \& Ramirez-Ruiz (2013)]{guillochon13} Guillochon J.,  \& Ramirez-Ruiz, E., 2013, ApJ, 767, 25

\bibitem[Guillochon et al.~(2014)]{guillochon14} Guillochon, J.,  Manukian, H.,  Ramirez-Ruiz, E., 2014, ApJ, 783, 23

\bibitem[Guillochon 
(2016)]{guillochon+magn} Guillochon, J. 2016, in prep

\bibitem[Haas et al.~(2012)]{haas+12} Haas, R., Shcherbakov, 
R.~V., Bode, T., \& Laguna, P.\ 2012, \apj, 749, 117 

\bibitem[Halpern et al.~(2004)]{halpern04} Halpern, J.~P., Gezari, S. \& Komossa, S. 2004, \apj, 604, 572

\bibitem[Hayasaki et al.~(2013)]{hayasaki13} Hayasaki, K., Stone, N., Loeb, A., 2013, MNRAS, 434, 909

\bibitem[Hayasaki et al.~(2015)]{hayasaki15} Hayasaki, K., Stone, N., Loeb, A., 2015, {\it submitted to MNRAS}

\bibitem[Hawley et al.~(1995)]{hawley+95} Hawley, J.~F., Gammie, 
C.~F., \& Balbus, S.~A.\ 1995, \apj, 440, 742 

\bibitem[Hawley et al.~(2011)]{hawley+11} Hawley, J.~F., Guan, X., 
\& Krolik, J.~H.\ 2011,  \apj, 738, 84 

\bibitem[Hawley et al.~(2013)]{hawley+13} Hawley, J.~F., Richers, S.~A., Guan, X., \& Krolik, J.~H.\ 2013, \apj, 772, 102 

\bibitem[Holoien et al.~(2014)]{holoien14} Holoien, T.~W.~S. et al. 2014, \mnras, 445, 3263

\bibitem[Holoien et al.~(2015)]{holoien15} Holoien, T.~W.~S. et al. 2015, \mnras, 455, 2918

\bibitem[Kobayashi et al.~(2004)]{kobayashi04a} Kobatayshi, S.,  Laguna, P., Phinney, E.S. \& Meszaros, P., 2004, ApJ, 615, 855

\bibitem[Komossa \& Bade~(1999)]{komossa99} Komossa, S. \& Bade, N. 1999, \aap, 343, 775

\bibitem[Komossa (2015)]{komossa15} Komossa, S., 2015, Journal of High Energy Astrophysics, 7, 147

\bibitem[Laguna et al.~(1993)]{laguna93b} Laguna, P.,  Miller, W.A., Zurek, W.H. \& Davies, M.B., 1993, ApJ, 410, L83

\bibitem[Lin et al.~(2015)]{lin15} Lin, D. et al. 2015, \apj, 811, 43

\bibitem[\protect\citeauthoryear{{Lucy}}{{Lucy}}{1977}]{lucy}
{Lucy} L.~B.,  1977, Astronomical Journal, 82, 1013

\bibitem[Maksym et al.~(2010)]{maksym10} Maksym, P., Ulmer, M.~P., Eracleous, M. 2010, \apj, 722, 1035

\bibitem[Maksym et al.~(2014)]{maksym14} Maksym, P., Lin, D. \& Irwin, J.~A. 2014, \apj, 792, L29

\bibitem[McKinney et al.(2012)]{jon+bigpaper} McKinney, J.~C., 
Tchekhovskoy, A., \& Blandford, R.~D.\ 2012, \mnras, 423, 3083 

\bibitem[\protect\citeauthoryear{{Monaghan}}{{Monaghan}}{2005}]{monaghan05}
{Monaghan} J.~J.,  2005, Reports on Progress in Physics, 68, 1703

\bibitem[\protect\citeauthoryear{{Morris} \& {Monaghan}}{{Morris} \&
  {Monaghan}}{1997}]{morris97}
{Morris} J.~P.,  {Monaghan} J.~J.,  1997, Journal of Computational Physics,
  136, 41
  

\bibitem[Narayan 
\& Yi(1994)]{narayanyi-94} Narayan, R., \& Yi, I.\ 1994, \apjl, 428, L13 


\bibitem[Novikov \& Thorne (1973)]{novikov73} Novikov, I.~.D. \& Thorne, K.~S. 1973, in \text{Black Holes}, ed. C.~DeWitt and B.~DeWitt (NY: Gordon \& Breach)

\bibitem[Papaloizou 
\& Pringle (1984)]{pp-84} Papaloizou, J.~C.~B., \& Pringle, J.~E.\ 1984, \mnras, 208, 721 


\bibitem[Penna et al.~(2013b)]{penna+alpha} Penna, R.~F., S{\c 
a}dowski, A., Kulkarni, A.~K., \& Narayan, R.\ 2013a, \mnras, 428, 2255 


\bibitem[Piran et al.~(2015a)]{piran+15a} Piran, T., S\k{a}dowski, 
A., \& Tchekhovskoy, A.\ 2015, \mnras, 453, 157 

\bibitem[Piran et al.~(2015b)]{piran+15b} Piran, T., Svirski, G., 
Krolik, J., Cheng, R.~M., \& Shiokawa, H.\ 2015, \apj, 806, 164 

\bibitem[Price (2012)]{price12a} Price 2012, Journal of Computational Physics, 231, 759

\bibitem[Rees (1988)]{rees88} Rees, M. J. 1988, Nature, 333, 523

\bibitem[Phinney (1989)]{phinney89} Phinney, E. S. 1989, IAU Symp. 136: The Center of the Galaxy, 543

\bibitem[Ramirez-Ruiz \& Rosswog (2009)]{ramirezruiz09} Ramirez-Ruiz, E., \& Rosswog, S.\ 2009, \apj, 679, L77 

\bibitem[\protect\citeauthoryear{{Rosswog00}, {Davies},
  {Thielemann} \& {Piran}}{{Rosswog} et~al.}{2000}]{rosswog00}
{Rosswog} S.,  {Davies} M.B., {Thielemann} F.-K., {Piran} T. 2000, 
A\&A, 360, 171

\bibitem[\protect\citeauthoryear{{Rosswog \& Price}}{2007}]{rosswog07}
{Rosswog} S. \& Price, D.J.,  2007, MNRAS 379, 915
 
\bibitem[\protect\citeauthoryear{{Rosswog09b}}{{Rosswog}}{2009}]{rosswogsph}
{Rosswog} S.,  2009, New Astronomy Reviews, 53, 78

\bibitem[\protect\citeauthoryear{{Rosswog08a}, {Ram\'irez-Ruiz} \&
  {Hix}}{{Rosswog} et~al.}{2008}]{rosswog08a}
{Rosswog} S.,  {Ram\'irez-Ruiz} E.,    {Hix} W.~R.,  2008, Astrophysical
  Journal, 679, 1385
  
\bibitem[\protect\citeauthoryear{{Rosswog08}, {Ram\'irez-Ruiz},
  {Hix} \& {Dan}}{{Rosswog} et~al.}{2008b}]{rosswog08b}
{Rosswog} S.,  {Ram\'irez-Ruiz} E.,    {Hix} W.~R., {Dan} M. 2008, 
Computer Physics Communications, 179, 184

\bibitem[\protect\citeauthoryear{{Rosswog09a}, {Ram\'irez-Ruiz} \&
  {Hix}}{{Rosswog} et~al.}{2009}]{rosswog09a}
{Rosswog} S.,  {Ram\'irez-Ruiz} E.,    {Hix} W.~R.,  2009, Astrophysical
  Journal, 695, 404

\bibitem[\protect\citeauthoryear{{Rosswog}}{2015c}]{rosswog15c}
{Rosswog} S.,  2015, Living Reviews of Computational Astrophysics, 1, 1

\bibitem[\protect\citeauthoryear{{Rosswog}}{2015b}]{rosswog15b}
{Rosswog} S.,  2015, MNRAS, 448, 3628

\bibitem[Roth et al.~(2015)]{roth15} Roth, N., Kasen, D., Guillochon, J. \& Ramirez-Ruiz, E. 2015,
arXiv:1510.08454

\bibitem[Saxton et al.~(2012)]{saxton12} Saxton, R.~D., {Read}, A.~M., {Esquej}, P., {Komossa}, S., 
	{Dougherty}, S., {Rodriguez-Pascual}, P. \& {Barrado}, D., 2012, \aap, 541, A106

\bibitem[S{\c a}dowski et al.~(2013)]{sadowski+koral} S{\c a}dowski, 
A., Narayan, R., Tchekhovskoy, A., \& Zhu, Y.\ 2013a, \mnras, 429, 3533 

\bibitem[S{\c a}dowski et al.~(2014)]{sadowski+koral2} S{\c a}dowski,
  A., Narayan, R., McKinney, J.~C., \& Tchekhovskoy, A.\ 2014, \mnras,
  439, 503 

\bibitem[S{\c a}dowski et al.~(2015a)]{sadowski+dynamo} S{\c a}dowski, 
A., Narayan, R., Tchekhovskoy, A., Abarca, D., Zhu, Y., \& McKinney
  J.~C. .\ 2015a, \mnras, 447, 49 


\bibitem[\protect\citeauthoryear{S\k{a}dowski \& Narayan}{S\k{a}dowski \&
  Narayan}{2015b}]{sadowski+3d}
S\k{a}dowski A.,  Narayan R. 2015d, MNRAS,
  in press 

\bibitem[Shiokawa et al.~(2015)]{shiokawa+15} Shiokawa, H., Krolik, 
J.~H., Cheng, R.~M., Piran, T., \& Noble, S.~C.\ 2015, \apj, 804, 85 

\bibitem[\protect\citeauthoryear{{Springel}}{{Springel}}{2010}]{springel10}
{Springel} V.,  2010, Annual Review of Astronomy and Astrophysics, 48,
391

\bibitem[Stone 
\& Metzger (2016)]{stone+16} Stone, N.~C., \& Metzger, B.~D.\ 2016, \mnras, 455, 859 


\bibitem[Tassoul(1978)]{tassoul-book} Tassoul, J.-L.\ 1978, 
Princeton Series in Astrophysics, Princeton: University Press, 1978,  


\bibitem[Tejeda 
\& Rosswog (2013)]{tejedarosswog-13} Tejeda, E., \& Rosswog, S.\ 2013, \mnras, 433, 1930 

\bibitem[Tejeda, Gafton, 
\& Rosswog (2015)]{TGR} Tejeda, E., Gafton, E., \& Rosswog, S.\ 2015, {\it in preparation}

\bibitem[Vink\'{o} et al.~(2014)]{vinko14} Vinkó, J. et al. 2014, \apj, 798, 12

\bibitem[van Welzen et al.~(2011)]{vanwelzen11} van Welzen, S. et al. 2011, \apj, 741, 73

\bibitem[Wielgus et al.~(2015)]{wielgus+mri} Wielgus, M., Fragile, 
P.~C., Wang, Z., \& Wilson, J.\ 2015, \mnras, 447, 3593 

\end{thebibliography}
}

\end{document}